\documentclass{elsarticle}
\makeatletter
\def\ps@pprintTitle{%
 \let\@oddhead\@empty
 \let\@evenhead\@empty
 \def\@oddfoot{}%
 \let\@evenfoot\@oddfoot}
\makeatother
\usepackage{lineno,hyperref}
\usepackage{amsmath,amssymb}
\usepackage{caption}
\usepackage[final]{changes}
\usepackage{graphicx}
\usepackage{xspace}
\usepackage{color,soul}
\usepackage{ulem}
\usepackage{natbib}
\usepackage{enumitem}
\newcommand{\Fg}[1]{Figure~{\ref{fig:#1}}}

\bibpunct{ (}{)}{;}{a}{,}{,}
\begin{document}

\begin{frontmatter}

\title{Spinning up planetary bodies by pebble accretion}

\author{R.G. Visser$^{a,\ast}$, C.W. Ormel$^a$, C. Dominik$^a$, S. Ida$^b$}

\address{$^{a}$Anton Pannekoek Institute, University of Amsterdam, Science Park 904, PO box 94249, Amsterdam, The Netherlands}
\address{$^{b}$Earth-Life Science Institute (ELSI), Tokyo Institute of Technology, Meguro, Tokyo, 152-8550, Japan}
\cortext[myfootnote]{Corresponding author. E-mail:  r.g.visser@uva.nl}

\begin{abstract}
\added[]{Most major planetary bodies in the solar system rotate in the same direction as their orbital motion: their spin is prograde. Theoretical studies to explain the direction as well as the magnitude of the spin vector have had mixed success. When the accreting building blocks are $\sim$ km-size planetesimals -- as predicted by the classical model -- the accretion process is so symmetric that it cancels out prograde with retrograde spin contributions, rendering the net spin minute.  For this reason, the currently-favored model for the origin of planetary rotation is the giant impact model, in which a single collision suffices to deliver a spin, which magnitude is close to the breakup rotation rate. However, the giant impact model does not naturally explain the preference for prograde spin. Similarly, an increasing number of spin-vector measurement of asteroids also shows that the spin vector of large (primordial) asteroids is not isotropic.  Here, we re-assess the viability of smaller particles to bestow planetary bodies with a net spin, focusing on the pebble accretion model in which gas drag and gravity join forces to accrete small particles at a large cross section. Similar to the classical calculation for planetesimals, we integrate the pebble equation of motion and measure the angular momentum transfer at impact. We consider a variety of disk conditions and pebble properties and conduct our calculations in the limits of 2D (planar) and 3D (homogeneous) pebble distributions. We find that in certain regions of the parameter space the angular momentum transfer is significant, much larger than with planetesimals and on par with or exceeding the current spin of planetary bodies. We link this large net spin delivery to the appearance of asymmetries during the accretion process of pebbles. For example, prograde contribution may dominate (in certain regions of the parameter space) because they originate from trajectories that are preferentially captured.  For simplicity, our calculations have ignored certain important effects (e.g., collisions, the back-reaction on the gas, and formation of atmospheres) and do not address how the eventual distribution of spin vectors is obtained for which collisions and post-formation processes must have played a role to explain the scatter.  Irrespective of these issues, pebble accretion is a viable mechanism to not only grow planetary bodies, but also to impart them with a significant spin.}
\end{abstract}

%Sketch of a secondary Hill sphere encounter of a pebble. At first approach the pebble escapes from the Hill sphere and drifts to the co-rotation line for which the shear (gray arrows) equals the headwind velocity (black arrow). As the pebble approaches the Hill sphere for a second time, the pebble is pulled inwards by the protoplanet gravity. During the path from $-R_\mathrm{Hill}$ to the protoplanet surface, the pebble is simultaneously accelerated downwards due to the negative shear velocity and headwind velocity. As a result the pebble is forced to impact counterclockwise below the protoplanet delivering prograde spin rotation.

\end{frontmatter}

\section{Introduction}

With the exception of Venus, all other planets in the solar system have their spin and orbital angular momentum vectors aligned (prograde rotation). The giant planets were likely spun-up by accretion through a circumplanetary disks\citep{Machida2008}. For planets dominated by solids -- the terrestrial planets as well as Uranus and Neptune -- the situation is more complex. The classical model of planet formation asserts that planetary cores grew by accretion of km-sized planetesimals \citep{PollackEtal1994}.  However, planetesimals do not provide planetary embryos with a significant amount of angular momentum\citep{IdaNakazawa1990,LissauerSafronov1991,DonesTremaine1993}. On average, the planar components do not bestow angular momentum by symmetry, $\langle l_x \rangle = \langle l_y \rangle =0$ where $\langle l_i \rangle$ is the average momentum per unit mass transferred by the impactor in the direction $i$. To obtain the vertical spin component $\langle l_z \rangle$ numerical integration of accreting planetesimal trajectories are conducted, averaging over impact parameter and phase angles. But these too display near-cancellation behavior; essentially, the point where planetesimals hit the surface of a planet is random, meaning that they are about as likely to impart positive angular momentum as negative angular momentum. In fact, the largest net contribution occurs for a dynamically cold planetesimal disk, for which the spin will be retrograde\citep{IdaNakazawa1990} (see \Fg{figure1}); the maximum prograde contribution (also indicated in \Fg{figure1}) occurs when the planetesimals are dynamically moderately excited, such that their scaleheight extends over a scale larger than the Hill radius of the body\citep{DonesTremaine1993}. This is more realistic, but it fails to reach the observed spins by many magnitudes. The only way for planetesimals to bestow a significant (prograde) spin is to invoke a non-uniform disk\citep{OhtsukiIda1998}, which may materialize when the planetary embryos grow in isolation, but not in a more general setting where multiple embryos scatter planetesimals around.

Because of the failure to deliver a \textit{systematic} spin by accreting planetesimals, the current paradigm holds that the primordial spin of planetary bodies was instead stochastically determined by late large impacts\citep{DonesTremaine1993i}. The spin would be stochastic when it is dominated by accretion of a single body that constituted a significant fraction of the target bodies' mass, as for example in the hypothesized formation of the Earth-Moon system. Indeed, the current paradigm for the formation of the terrestrial planets asserts the existence of a giant impact phase, where many Mars-size embryos were perturbed onto crossing orbits after the dispersal of the gaseous disks\citep{Wetherill1985}. These equal-mass embryos collide at velocities similar to the surface escape velocity, conveying an angular momentum close to the breakup frequency $\omega_\mathrm{crit}=\sqrt{GM_p/R^3}$ \citep{KokuboIda2007,MiguelBrunini2010}. Importantly, the giant impact scenario also predicts that the distribution of the spin vector must be isotropic, such that the spin axis is oriented towards low ecliptic latitude, as is the case with Uranus but not with the terrestrial planets.  However, this result is not considered a problem for the giant impact hypothesis, as the number of planets is limited and the spin of the terrestrial planets is known to be significantly affected by post-formation processes \citep{Dobrovolskis1980}. For example, the spin obliquity for Mars is known to be unstable (chaotic) on timescales short as $\sim$10 Myr \citep{LaskarRobutel1993}. 

The asteroid belt provides us with an opportunity to inform us on the primordial spin state of solar system bodies, as the spin vector of a growing number of bodies is measured\citep{WarnerEtal2009}. Small asteroids are believed to be collisional products, which have erased any memory of the primordial spin. In addition, the spin of small asteroids is affected by post-formation processes like YORP \citep{Rubincam2000}. On the other hand, the spins of asteroids with diameter $D>120$ km are thought to have been hardly affected by collisions or post-formation dynamics
\citep{Bottkeetal2005,SteinbergSari2015}. In \Fg{figure1} we plot the spin vector (in terms of its rotation frequency and ecliptic latitude $\beta$) for asteroids of $D>150$ km. Although it is obvious that asteroids evolved in a highly collisional environment, \Fg{figure1} clearly displays a preference for a prograde spin compared to what one would expect from an isotropic distribution. A Kolmogorov-Smirnov test indicates that \added[]{the null-hypothesis of a uniform distribution in $\sin \beta$ can be rejected with a probability of $>$ 99\% }. This result hints that asteroids assembled through a mechanism that provided them with a systematic (vertical) spin. Similarly to the asteroid belt, recent observations of the mutual orbit spin orientations of trans-neptunian binaries also show a preference for prograde motion \citep{Grundyetal2019} suggesting that prograde spin is the natural starting state of (solar system) planetary bodies.

Pebble accretion\citep{OrmelKlahr2010,LambrechtsJohansen2012} is an alternative theory to explain the assembly of planetary bodies. In pebble accretion bodies accrete aerodynamically small particles (``pebbles'') which drift inwards through the disk by virtue of the sub-Keplerian motion of the gas \citep{Weidenschilling1977}. The pebbles are then captured by the combined effects of gravity and gas drag to spiral inwards (settle) towards the protoplanet surface. Pebble accretion has been invoked to help explain the architecture of the solar system: its terrestrial planets \citep{LevisonEtal2015}, its giant planets \citep{LevisonEtal2015i} and its asteroids\citep{JohansenEtal2015}.

\added[]{The accumulation of planetary spin by pebble accretion has not been deeply studied. \citet{JohansenLacerda2010} were the first to investigate the link between accreting small particles and the spin of the accreting body in a gaseous environment. The authors performed local hydrodynamical simulations (a shearing sheet with periodic boundary conditions), involving high pebble densities resulting in instantaneous and very rapid accretion. Because of the high densities particle feedback effects are important, resulting in a prograde spinning circumplanetary disk. While these pioneering simulations are insightful in highlighting the ability of small particles to spin up bodies, they rely on rather specific conditions. \\
In this paper we explore the implications pebble accretion has for the spin of accreting bodies in a more general fashion. We calculate the angular momentum that is being transfered to the accreting body by integrating the equation of motion of individual pebbles. Summing over the range of impact trajectories we find, in this way, the net AM that is being transfered to the accreting body. Although neglecting feedback effects, our  more "basic" approach has the advantage that we can easily scan a wide parameter space and in this way identify specific trends in the simulated data.
Our approach is by-and-large similar to the classical calculations for planetesimals \citep{IdaNakazawa1990,LissauerSafronov1991,DonesTremaine1993} but then tailored towards pebbles. Specifically, we will show in this paper that spin transfer is connected to the appearance and disappearance of asymmetries in the pebble accretion process. We show that these asymmetries lead to mean spin outcomes which are on a par with what is observed in the asteroid belt and planets.}

The paper is structured as follows. In Section \ref{sec:nummod} we discuss our numerical model, in Section \ref{sec:results} we present the obtained results, the discussion of the results is given in Section \ref{sec:Discussion} and we relate our results to the solar system in Section \ref{sec:SSB}. The key conclusions are presented in Section \ref{sec:concl}. 

\section{Numerical model}
\label{sec:nummod}
To determine the amount of rotation pebbles supply to the accreting protoplanet at impact, pebble trajectories are numerically integrated \added[]{with a Runge-Kutta-Fehlberg variable step scheme (RKF45) \citep{Fehlberg1969}, (used by \citet{VisserOrmel2016}).} We use a reference frame co-moving with a protoplanet with mass $M_p$ at distance $r_0$ from the star (Figure \ref{fig:sketch}). The three dimensional equation of motion (EOM) for a test particle with velocity $\mathbf{v} = (v_x,v_y,v_z)$ and position $\mathbf{r} = (x,y,z)$ in the local frame is given by \citep{IdaNakazawa1990,OrmelKlahr2010}:
\begin{equation}
    \frac{\mathrm{d}\mathbf{v}}{\mathrm{d}t} = 
    \begin{pmatrix}
         2\Omega_0 v_y + 3\Omega_0^2x\\ 
        -2\Omega_0v_x\\ 
         0
    \end{pmatrix} 
    -\frac{GM_p}{r^{3}}
    \begin{pmatrix}
        x\\ 
        y\\ 
        z
    \end{pmatrix} 
	+\mathbf{F}_\mathrm{d},
\label{eq:eqofmotion}
\end{equation}
with $G$ the universal gravity constant, $(2 \Omega_0 v_y, -2 \Omega_0 v_x, 0)$ the Coriolis acceleration, $3\Omega_0^2x$ the tidal force and $\Omega_0$ the Keplerian frequency around the star. The second and last term contain the protoplanet gravity and gas drag, respectively. The stellar gravity component in the $z$ dimension has been omitted, an approximation that is further motivated in Section \ref{sec:3d}. The gas drag force is given by:
\begin{equation}
\mathbf{F}_\mathrm{d}= -\frac{\mathbf{v} - \mathbf{v}_\mathrm{g}}{t_s},
\end{equation}
with $t_s$ the stopping time, the time after which solids are slowed down by the gas. For pebbles of radius $s$ and internal density $\rho_{\bullet\mathrm{s}}$ \citep{Whipple1972}:
\begin{equation}
    t_s = \left\{\begin{matrix} \displaystyle \frac{\rho_{\bullet\mathrm{s}} s}{\rho_{\mathrm{g}} v_{\mathrm{th}}} &                \textrm{Epstein regime: }  \quad s<\frac{9}{4}l_{\mathrm{mfp}} \\[5mm] 
    \displaystyle
    \frac{2 \rho_{\bullet\mathrm{s}}s^{2}}{9 \eta_d} & \textrm{Stokes regime:} \quad s \geq \frac{9}{4}l_{\mathrm{mfp}}\\ 
\end{matrix}\right.,
\label{eq:tstop} 
\end{equation}
where the Stokes regime is valid for low Reynolds number, Re $\ll$ 1, with $\eta_d$ the dynamic viscosity, $l_{\mathrm{mfp}}$ the molecular mean free path, $v_{\mathrm{th}}$ the thermal speed and $\rho_{\mathrm{g}}$ the gas density. The gas flow is approximated as the Keplerian shear motion around the planet:
\begin{equation}
\mathbf{v}_\mathrm{g} = (-v_\mathrm{hw} - \frac{3}{2}\Omega_0 x)\hat{\mathbf{y}},
\label{eq:vgas}
\end{equation} 
with $v_\mathrm{hw}$ the gas headwind which arises because the gas orbits the star with a speed slightly lower than the Keplerian speed $v_k$ due to pressure support pointing radially outwards \citep{Weidenschilling1977}. A large body in an unperturbed Keplerian orbit, in a co-moving reference frame, hence faces a headwind of magnitude:
\begin{equation}
    v_\mathrm{hw} = \eta v_k ,
\label{eq:headwind}
\end{equation} 
where $\eta$ measures the deviation from the Kepler velocity \citep{Nakagawa1986}:
\begin{equation}
    \eta = \frac{c_s^2}{2v_k}\frac{\partial \log P}{\partial \log r_0 } \sim \left(\frac{c_s}{v_k} \right)^2,
    \label{eq:eta}
\end{equation}
and $P$ is the gas pressure in the midplane.

The Hill radius of the accreting body is given by:
\begin{equation}
    R_\mathrm{H} = r_0\left ( \frac{M_p}{3M_\star } \right )^{1/3},
    \label{eq:hill}  
\end{equation} 
with $M_\star$ the mass of the central star. Normalizing the velocities to the Hill velocity $R_\mathrm{H} \Omega_0$, distances to the Hill radius $R_\mathrm{H}$ and units of time to the inverse orbital frequency of the protoplanet $\Omega_0^{-1}$ in line with \citet{OrmelKlahr2010}, the 3D pebble EOM in Hill units is given by:
\begin{equation}
    \frac{\mathrm{d} }{\mathrm{d} t}
\begin{pmatrix}
v_x \\ 
v_y\\
v_z
\end{pmatrix}
=\begin{pmatrix}
2v_y + 3x-3x/r^3\\ 
-2v_x-3y/r^3\\
-3z/r^3
\end{pmatrix}-\frac{1}{\tau_s}
\begin{pmatrix}
v_x \\ 
v_y+ \zeta_w + 3x/2\\
v_z
\end{pmatrix},
\label{eq:dimlessEOM}
\end{equation} 
velocities and positions are now represented as dimensionless quantities, with $GM_p = 3$ in Hill units. The dimensionless headwind parameter $\zeta_w$ is given by \citep{OrmelKlahr2010}:
\begin{equation}
    \zeta_\mathrm{w} \equiv \frac{v_\mathrm{hw}}{\Omega_0 R_\mathrm{H}},
    \label{eq:zeta}
\end{equation} 
which is therefore an indicator of the  mass (gravity) of the protoplanet. The physical protoplanet radius is normalized by the Hill radius:
\begin{equation}
    \alpha_p \equiv \frac{R}{R_\mathrm{H}}.
    \label{alpha1}
\end{equation}
and the Stokes number is given by:
\begin{equation}
\tau_s = t_s \Omega_0.
\end{equation}

In these units, the input parameters for our integrations are reduced to $\zeta_w$, $\alpha_p$ and pebble Stokes number $\tau_s$. A direct relation between the main physical and dimensionless quantities is summarized in Table \ref{tab:tabledimvsdimless}. 

In general specific values of disk parameters are not known and vary greatly since protoplanetary disks are subject to complex time evolution during their lifetime (e.g.\citep{Bitsch2015}). Nevertheless, we will occasionally transform our results into physical quantities adopting power-law profiles for both the gas temperature and gas surface density profile following from the the Minimum Mass Solar Nebula (MMSN)\citep{Weidenschilling1977B,HayashiEtal1985,Nakagawa1986}:
\begin{equation}
T(r) = 170\ \mathrm{K}\left ( \frac{r_0}{1 \ \mathrm{au}} \right )^{-1/2},
\label{eq:Temp}
\end{equation}
\begin{equation}
\Sigma(r) =1300 \ \mathrm{g \ cm^{-2}}\left ( \frac{r_0}{1 \ \mathrm{au}} \right )^{-3/2}.
\label{eq:surfdens}
\end{equation}
\deleted[]{The assumed power law profiles are slightly adjusted with respect to the Minimum Mass Solar Nebula (MMSN) to account for more efficient cooling.}
 The disk gas scaleheight is given by:
\begin{equation}
    H_\mathrm{g}=\frac{c_s}{\Omega_0}.
    \label{eq:gasscaleheight}
\end{equation}
Assuming an isothermal column, the gas distribution in the vertical dimension $z$ is equal to:
\begin{equation}
    \rho_\mathrm{g} =\frac{\Sigma }{H_\mathrm{g}\sqrt{2\pi}} \exp\left[{-\frac{1}{2}\left ( \frac{z}{H_\mathrm{g}} \right )^2}\right].
    \label{eq:gasdens}
\end{equation}
The main input parameters $\zeta_w$, $\alpha_p$ and pebble Stokes number $\tau_s$ can now be expressed in terms of the disk parameters as:
\begin{flalign}
 &v_\mathrm{hw} = 30 \ \mathrm{ms^{-1}} \left ( \frac{\mu}{2.34} \right )^{-1} \left (\frac{M_\star}{M_\odot } \right )^{-1/2},& \\
    &\zeta_\mathrm{w} = 12.5\left ( \frac{\rho_\bullet}{2.5 \ \mathrm{g \ cm^{-3}}} \right )^{-1/3}\left ( \frac{v_\mathrm{hw}}{30 \ \mathrm{m \ s^{-1}}} \right )\left ( \frac{R}{100 \ \mathrm{km}} \right )^{-1}\left ( \frac{r_0}{\mathrm{au}} \right )^{1/2},& \\
    &\alpha_p = 5.7 \times 10^{-3} \left ( \frac{M_\star}{M_\odot } \right )^{1/3}\left ( \frac{\rho_\bullet}{2.5 \ \mathrm{g \ cm^{-3}}} \right )^{-1/3}\left ( \frac{r_0}{\mathrm{au}} \right )^{-1}, & \\ 
&\tau_s = 0.1 \left ( \frac{s}{1 \ \mathrm{cm}} \right ) \left ( \frac{\rho_{\bullet s }}{2.5 \ \mathrm{g\ cm^{-3}}} \right )\left ( \frac{M_\star}{M_\odot } \right )^{1/2}\left ( \frac{r_0}{\mathrm{au}} \right )^{-3/2},
\end{flalign} 
with $\mu$ the mean molecular weight in atomic mass units and where we applied the ideal gas law.

\subsection{Pebble-to-planet spin supply}
The spin supply of pebbles to the protoplanet is calculated by tracing individual specific angular momentum (SAM) contributions from pebbles at impact on the protoplanet.
An individual pebble heading for the protoplanet bestows a SAM of:
\begin{equation}
 \mathbf{l} = \mathbf{r}\times \mathbf{v}.
\end{equation} 
The $z$-component is the one of interest:\begin{equation}
l_{z} = \left ( xv_y - yv_x \right ) + \Omega_0\left ( x^2 + y^2 \right ),
\end{equation} where the latter term is to correct for the rotation of the co-moving coordinate frame \citep{DonesTremaine1993}.
The mean SAM transferred to the protoplanet is found by averaging the SAM contributions over the accretion cross-section. For example, in 2D:
\begin{equation}
    \left \langle l_z \right \rangle = \frac{\int_{\mathrm{accret}}F(x_0)l_z(x_0)\mathrm{d}x_0 }{\int_{\mathrm{accret}} F(x_0)\mathrm{d}x_0},
    \label{eq:meanspin}
\end{equation}
with $F(x_0) = v_{y,\infty} \Sigma $ the flux of pebbles entering our domain and $\Sigma$ the pebble surface density. It is useful to normalize the specific mean spin angular momentum $\left \langle l_z \right \rangle$ with $l_\mathrm{z,esc} = \sqrt{2 G M R}$ for the analysis of the results. Converting this to a fraction of the break-up (critical) frequency gives:

\begin{equation}
    \frac{\omega_z}{\omega_\mathrm{crit}}\approx 3.5 \frac{\left \langle l_z \right \rangle}{l_\mathrm{z,esc}},
\end{equation}
with $\omega_\mathrm{crit} = \sqrt{G M / R^3}$.

\subsubsection{Initial conditions}
Integration of pebble trajectories start at a position $\boldsymbol{x}=(x_0,y_0,z_0)$ (see Figure \ref{fig:sketch}). We take $y_0$ far from the protoplanet with radius $R$, to ensure that the planet gravity is initially insignificant. Pebbles are initiated with the unperturbed azimuthal and radial drift velocities \citep{Weidenschilling1977}:
\begin{align}
    &v_{x,\infty} =-\frac{2v_\mathrm{hw}\tau_s}{1 + \tau_s^2}, & \\
    &v_{y,\infty} = -\frac{v_\mathrm{hw}}{1 + \tau_s^2} -\frac{3}{2} \Omega_0 x_0.
    \label{eq:vxin}
\end{align}
The vertical release velocity $v_{z,\infty} = 0$. \added[]{To ensure that the pebbles are initially not influenced by the protoplanet gravity we demand that the gas drag force is larger than the two-body gravitational force of the protoplanet by a safety factor C. Specifically, we fix $y_0 = C\sqrt{GMt_s / v_\mathrm{hw}}$, with $C = 100$ \citep{OrmelKlahr2010}. }
\deleted{We fix $y_0 = 100\sqrt{GMt_s / v_\mathrm{hw}}$ to ensure an unperturbed release distance.} We then scan along the $y=y_0$ line to determine for which range in $x_0$ pebbles are accreted. The condition for accretion of a pebble onto the protoplanet is given by:
\begin{equation}
r -R \leq 0.
\end{equation}
In this way we find $x_1$ and $x_2$, the interior, respectively, exterior edge of the protoplanet impact range.

In the 3D case the collision range is symmetric around $z=0$ and we release pebbles over an interval $z_0 \in [-b_z,b_z]$ while ensuring that misses are included on both sides of the impact range.

\section{Results}
\label{sec:results}

The main result of this study are highlighted in Section \ref{sec:spincont}, which summarizes the result of our numerical integrations (\Fg{figure2}). In Section \ref{sup:analysis} (2D) and Section \ref{sec:3d} (3D) we provide an in-depth trajectory analysis, which allows us to understand the trends that we observe in the simulated data. We present test cases with earlier work in Section \ref{sec:DT}.

\subsection{Spin contributions}
\label{sec:spincont}
In \Fg{figure2} the imparted spin as function of pebble stopping time $\tau_s$ and headwind-to-shear parameter $\zeta_w$ is plotted for $\alpha_p=2\times10^{-3}$, which corresponds to a distance of about 2.5 au from a solar-mass star and a planet density of 2.5 $\mathrm{g \ cm^{-3}}$. The top panel of Figure \ref{fig:figure2} provides $\omega_z/\omega_\mathrm{crit}$ for a planar configuration (pebbles having settled to a midplane), whereas the bottom panel gives $\omega_z/\omega_\mathrm{crit}$ in the 3D limit, applicable when the pebble scaleheight exceeds the impact radius for pebble accretion. The 3D configuration is therefore more applicable for small pebbles that experience strong turbulence and for small planets. These results show that large prograde spin can be obtained even beyond the levels seen in the solar system's bodies. For example, the spin of Ceres and other asteroids ($\omega/\omega_\mathrm{crit}\sim0.3$; thick contour) is well matched by accretion of $\tau_s\approx0.05$ pebbles either in the 2D or 3D mode.  The spin of smaller asteroids ($R\sim100$ km) can also be attributed to pebble accretion, provided that the headwind velocity is less than $30\,\mathrm{m\,s}^{-1}$. Retrograde spins are also possible, albeit in small regions of the parameter space. \Fg{figure2}b shows 3D accretion of larger pebbles may spin up planets up to their breakup rate. Values $\omega_z>\omega_\mathrm{crit}$ are however artificial; in reality the pebble will bounce off to be decelerated by gas drag until it is re-accreted at the critical spin frequency, resulting in these bodies to rotate at breakup speeds. In any case, it is more likely that in those cases the interaction geometry follows the 2D limit. 

Pebble accretion is in fact quite efficient to deliver a systematic spin. The degree of asymmetry between prograde and retrograde spin contributions, $\delta = \langle l_z \rangle / \sqrt{\langle l_z^2 \rangle}$ exceeds $\delta = 0.1$ over a large part of the parameter space (see \Fg{delta}), whereas for planetesimal accretion $\delta\sim\alpha_p\sim10^{-3}$\citep{DonesTremaine1993}, see Section \ref{sec:DT}.  A qualitative understanding of this result follows from considering asymmetries in the interaction timescale of pebbles that encounter the planet on an interior vs exterior trajectory, see \Fg{figure3}. Pebble accretion first starts to operate on the exterior trajectories, which are flung back in a direction $\approx$45$^\circ$ against the disk headwind due to the drift angle. Hence these encounters last long, resulting in their capture, while the interior trajectories momentarily avoid accretion. This explains the prograde band at $R=R_\mathrm{PA}$, the radius where pebble accretion commences \citep{VisserOrmel2016}. For larger planets, on the other hand, encounters last long for the pebbles entering the Hill sphere on interior orbits due to the importance of the shear motion (\Fg{figure3}b). This again results in a prograde net contribution, which is even stronger in the 3D (see \Fg{figure3}b and Section \ref{sec:3d}) 
\subsection{Trajectory analysis for $\tau_s=0.1$ pebbles \label{sup:analysis}}
In the co-moving frame we classify pebbles that cross the x-axis initially from $x\lesssim -R$ as \textbf{interior trajectories} and pebbles that cross the x-axis initially from $x\gtrsim R$ as \textbf{exterior trajectories}. There are three directions in which spin can be imparted to the protoplanet by a pebble:

\begin{enumerate}
    \item pebbles fall in counter-clockwise to the protoplanet imparting spin aligned with the orbital rotation $\Omega_0$ (prograde)
    \item pebbles fall in perpendicular to the protoplanet surface, imparting no net spin 
	\item pebbles fall in clockwise to the protoplanet imparting spin opposite to the orbital rotation $\Omega_0$ (retrograde) 
	
\end{enumerate}

In the case of symmetry the sum of individual pebble spin contributions over the collision cross-section is zero. However, as it turns out several asymmetries produce non-zero mean spin outcomes.

We use the example model for fixed $\tau_s =0.1$ and $\alpha_p = 0.003$ (Fig.\ref{fig:lzgraph}, blue curve) to analyze pebble trajectories. We select five protoplanet radii from this curve corresponding to the most extreme transition between prograde and retrograde outcomes throughout the curve. For each protoplanet radius we then select several pebble trajectories ranging from an interior miss, a range of impacts and an exterior miss (Fig. \ref{fig:ballpebb}a to e, left panel). The right panel of Figure \ref{fig:ballpebb}a to e shows the specific angular momentum an individual pebble imparts at impact on the protoplanet (every red dot), with respect to the release distance $x_0$ within the collision cross-section. The impacting trajectories in the left panel correspond to the highlighted dots with equal color in the right panel, going from dark (most interior trajectory) to light (most exterior trajectory). The highlighted dots in the right panel are uniquely labelled for reference in the text. 

As protoplanet radii in our simulations range from $R$= 10 km to $R$=3000 km there are a range of different accretion regimes which we will use to structure the analysis. These regimes are also shown in Figure \ref{fig:delta} where we show the degree of asymmetry between prograde and retrogade contributions for the 2D case (blue curve). Specifically, for our parameter space the following regimes are relevant:

\begin{enumerate}[label=(\alph*)]
    \item geometrical and ballistic accretion\citep{Safronov1972}, $R\lesssim 200\ \mathrm{km}$(Fig. \ref{fig:ballpebb}a)
    \item onset pebble accretion \citep{VisserOrmel2016}, $ 200\ \mathrm{km} \lesssim R \lesssim 400\ \mathrm{km}$(Fig. \ref{fig:ballpebb}b)
\item pebble accretion\citep{OrmelKlahr2010,LambrechtsJohansen2012}, $ 400\ \mathrm{km} \lesssim R \lesssim 1000\ \mathrm{km}$(Fig. \ref{fig:ballpebb}c)
\item transition to shear dominated pebble accretion, $ R \sim 1000\ \mathrm{km}$(Fig. \ref{fig:ballpebb}d)
    \item shear dominated pebble accretion \citep{OrmelKlahr2010,LambrechtsJohansen2012} $ R \gtrsim 1400\ \mathrm{km}$(Fig. \ref{fig:ballpebb}e)
\end{enumerate}

\subsubsection*{Regime (a): Geometrical accretion and ballistic accretion} 
For both geometrical and ballistic impacts, prograde pebble spin imparted by impacts from the interior band (Fig. \ref{fig:ballpebb}a, trajectories between a0 and a1) is canceled by the exterior band ((Fig. \ref{fig:ballpebb}a, trajectories between a1 and a2), which deliver retrograde spin. The sum of all the spin contributions sum to zero and symmetry prevails. 

\subsubsection*{Regime (b): Onset pebble accretion}
Pebble accretion commmences at protoplanet radius \citep{VisserOrmel2016}:
\begin{equation}
    R_\mathrm{PA}
\approx 160\ \mathrm{km}\
    \left( \frac{v_\mathrm{hw}}{\mathrm{30\ m\ s^{-1}}} \right)
    \left( \frac{\rho_\bullet}{\mathrm{2.5 \ g\ cm^{-3}}} \right)^{-0.36}
    \left( \frac{r_0}{\mathrm{au}} \right)^{0.42}  \left( \frac{\tau_s}{0.1} \right)^{0.28},
    \label{eq:onsetpa}
\end{equation}
which is valid for $\tau_s < 1$. From $R_\mathrm{PA}$, pebbles at interior trajectories ((Fig. \ref{fig:ballpebb}b, trajectories between b0 and b1) start to differ from pebbles at exterior trajectories((Fig. \ref{fig:ballpebb}b, trajectories between b1 and b3). In particular trajectory b0 is still accreting in a ballistic fashion equivalent to trajectory a0. The exterior trajectories on the other hand (b2 and b3) are captured last moment already entering the settling regime, enhancing the exterior accretion band.
That (axial) symmetry is broken can also be seen from two trajectories that both reach the same minimum distance $r_\mathrm{min}$ upon first approach to the protoplanet. The interior blue trajectory escapes, while the red exterior trajectory hits the protoplanet with a prograde contribution (Fig. \ref{fig:figure3}). This asymmetry between interior and exterior encounters can be understood from the angle at which the drifting pebbles approach the protoplanet. For this reason the escaping interior trajectory is deflected more \textit{downwards} and trajectory b2 is deflected more \textit{upwards}. Consequently, encounter (crossing) times are longer for trajectory b2, resulting in its capture. 

Naively it would be expected that the exterior trajectories b2 and b3 accrete clockwise (retrograde) by chasing spirals. Instead they fly across the protoplanet again, reverse orbit and wrap around the planet from below to impart a significant amount of counter-clockwise (prograde) spin. Effectively it can be conluded that the exterior band already experiences a transition from ballistic encounters (too fast to be captured) to settling encounters in which the protoplanet gravity is able to capture the pebble. Because this is the transition to settling, pebbles are captured at the verge of escape, leading to the last moment orbit reversal back to the protoplanet surface, to impart prograde spin. The onset of pebble accretion stated in Equation \ref{eq:onsetpa} is therefore essentially only valid for the exterior band. At somewhat larger protoplanet radius (gravity) $R \gtrsim 300$ km, the interior band catches up by also accreting pebbles through the settling mechanism, equivalent to the exterior band between b2 and b3. The interior band imparts spin in a retrograde fashion, restoring the symmetry and bringing mean spin  $ \langle l_z \rangle$  close to zero again (Fig. \ref{fig:ballpebb}b-right panel).

\subsubsection*{Regime (c): Pebble accretion}
For $R = 400$ km pebble trajectories mean spin is dominated by retrogade contributions due to an asymmetry associated with the in-spiralling process. Trajectories c1 and c3 are captured and spiral inwards due to settling. However, before even finishing half a spiral these pebbles already collide with the protoplanet surface. (Fig. \ref{fig:ballpebb}c-left and right panel). Again the interior trajectory c1 is deflected more downwards than the exterior trajectory c3. Consequently, trajectory c1 is prevented from delivering a large amount of prograde rotation since it is guided toward a perpendicular impact to the protoplanet surface. Trajectory c3 on the other hand is able to wrap around the protoplanet, finally "grazing" the protoplanet surface at impact to impart significant retrograde spin. With increasing gravity the symmetry is again restored for $R \sim 500$ km as pebbles completely spiral in (settle) (like the c0 and c4 trajectories) for both the interior and exterior bands.

\subsubsection*{Regime (d): Transition to shear dominated pebble accretion}
For protoplanet radius $R \sim 1000$ km the background Keplerian shear velocity ($v_\mathrm{sh} = -1.5 \Omega_0 x$) becomes comparable to the gas headwind velocity $v_\mathrm{hw}$ since collision cross-sections reach a significant fraction ($\sim 0.5R_\mathrm{H}$) of the protoplanet Hill sphere. The $x-$coordinate for which the gas headwind velocity equals the upwards Keplerian shear layer is given by:
\begin{equation}
    x_\mathrm{co} = \frac{2}{3} \Omega^{-1} v_\mathrm{hw},
\end{equation} 
which is referred to as the co-rotation line. As pebbles drift closer to this line, their y-velocity becomes decreasingly negative to eventually follow horseshoes upwards (gray missing trajectories in Fig. \ref{fig:ballpebb}d-left panel).

Interior pebbles experience longer encounter times as they approach the co-rotation line while drifting inwards. As long as $x_\mathrm{co} < -R_\mathrm{H}$ this enhances the interior accretion (the band between trajectory d0 and d2 in Fig. \ref{fig:ballpebb}d) since gravity has a long time to capture these pebbles. These pebbles accrete prograde because they fall in with counter-clockwise spirals. Exterior pebbles on the other hand have a more reduced encounter time; they approach in shear layers with a higher downwards velocity (the band between trajectory d2 and d4 in Fig. \ref{fig:ballpebb}d). As a consequence, the interior band (between d0 and d2) is visibly broader than the exterior one (between d2 and d4) as shown in the right panel of Fig. \ref{fig:ballpebb}d. Since the interior pebbles spiral in counter-clockwise to the protoplanet, the mean spin outcome is prograde. Thus, the transition to the shear-regime implies another asymmetry, which favors prograde contribution.

\subsubsection*{Regime (e): Shear dominated pebble accretion}
For $R \sim 3000$ km the Hill radius extends beyond the co-rotation line (Fig. \ref{fig:ballpebb}e, left panel); collision cross sections are now entirely determined by the Keplerian shear. Although a small part of the interior band (Fig. \ref{fig:ballpebb}e, trajectories e0 to e1) still accretes in a prograde fashion a significant part has been lost after crossing the co-rotation line to follow upwards horseshoes (interior gray miss, Fig. \ref{fig:ballpebb}e, left panel). Consequently, the retrograde contribution from exterior trajectories dominate. 

A subset of the exterior trajectories, however, now accrete from the back. After having traveled through the Hill sphere once, they cross the co-rotation line due to radial drift. The pebble then travels upwards from $-y$ and enters the Hill sphere a second time (trajectory e4, Fig. \ref{fig:ballpebb}e, left panel). These back-accreting pebbles will always supply prograde angular momentum, as we can see easily in the following way: The protoplanet orbits with the Keplerian velocity while the pebble is pulled from $x\sim x_\mathrm{co}$ to $x = R$, starting with $v_y\sim 0 $ at the co-rotation line and then, while approaching the planet, moving toward the downward headwind speed and the downwards shear field combined. This forces the pebble to impact the protoplanet in a counter-clock-wise fashion resulting in a maximum prograde contribution to the angular momentum  (see Figure \ref{fig:Horseshoe} for clarification). Nonetheless, the total prograde contributions do not outweigh the retrograde ones and the mean spin outcome is dominantly retrograde. 

For even higher protoplanet gravity the Hill sphere extends even further beyond the co-rotation line making the secondary contributions such as trajectory e4 take over the full accretion band. As a consequence mean spin outcomes are effectively prograde again as shown in Figure \ref{fig:lzgraph} for $R \gtrsim 4000$ km.
\color{black}

\subsection{Analysis of the 3D simulations}
\label{sec:3d}
Generally, pebbles are expected to reside in a sub-disk, in which the pebble sedimention is balanced by turbulent diffusion \citep{Dubrulleetal1995}. The thickness of the particle disk is therefore determined by the aerodynamical properties of the pebbles ($\tau_s$) as well as the strength of the turbulence (usually characterized in terms of a diffusivity $D=\alpha_z c_s H$). Therefore the pebbles should be modelled according to a Langevin equation \citep{OrmelLiu2018} or with hydrodynamical simulations \citep{Homannetal2016}. We opt to avoid these formal but complicated models by omitting the vertical \textit{stellar} gravity component. In the absence of the protoplanet gravity, pebbles would then preserve their height ($z$). In addition, in our 3D calculations we assume an homogeneous distribution of pebbles in $z$, i.e., we assume that the pebble scale height is significantly larger than the Hill radius of the protoplanet. \added[]{In the 3D case turbulent trajectories will be ballistic ("straight") over timescales less than the turbulence correlation time. Since the latter is typically $\sim\Omega^{-1}$ in most turbulence models \citet{Cuzzietal2001}, longer than the settling timescale, 3D accretion onto small planets/asteroids can be approximated as ballistic. Then, our 2D and 3D simulations constitute the two limits of the general situation.}

In Figure \ref{fig:lzgraph3d} 3D simulations for $\alpha_p = 3 \ \times \ 10^{-3}$ and for $\tau_s =0.1$ show that mean spins are dominantly more prograde in the 3D case with respect to the 2D case. We provide a similar analysis as applied to the 2D results for the most notable transitions in the 3D spin curves shown in Figure \ref{fig:lzgraph}. We will show that most of the features in the 2D case can be extended to the 3D case. Heat maps of individual spin contributions ${l}_{z,i}$ at impact are shown in Figure \ref{fig:indcont3d} for initial release distance $z_0$ on the y-axis and initial release distance $x_0$ on the x-axis. 

The physical processes responsible for the behavior of the 2D spin curve in section \ref{sup:analysis} are equivalent in 3D. The trajectory types in the 2D case described in Figure \ref{fig:ballpebb} apply to a mid-plane slice of these heat maps ($z_0 =0$). The trajectory types are however non-changing for non-zero release distance $z_0$, while keeping $x_0$ fixed. This is indicated in Figure \ref{fig:indcont3d} with the vertical solid lines with the corresponding unique label shown in the 2D case (Fig. \ref{fig:ballpebb}).    

The main difference in 3D, favouring the prograde spin, is the increase in the encounter time discrepancy between interior and exterior pebbles due to shear. In the shear important regime interior pebbles released at $(x_0,y_0,z=0)$ drift to the co-rotation line (trajectory e0, for example), gaining small velocities and consequently long encounters. This effect remains unchanged in 3D since the influence of Keplerian shear and drift on a pebble are equivalent for release distance $(x_0,y_0,z \neq 0)$ (Fig. \ref{fig:indcont3d}\textbf{D}, trajectory e0). The long lasting interior encounters for $z_0 \neq 0$ lead to timely vertical settling to the mid-plane and a strong $l_z$ contribution. In particular the following effects are observed in 3D:

\begin{enumerate}
	\item Analogous to the 2D results, the mean spin is effectively zero for the geometrical and ballistic regime(Fig. \ref{fig:indcont3d} \textbf{A}).
	\item A larger prograde contribution compared to the 2D case due to the transition from a 1D to a 2D collision cross-section (larger surface ratio, Fig. \ref{fig:indcont3d} \textbf{B}),
    \item Shear dominated regime; comparatively longer encounter times for interior pebbles (trajectory d0 to d2 and e0 to e1), which are accreted accross a larger $z$. Vertical settling is made possible for interior pebbles since they drift closer towards the co-rotation line, promoting long lasting encounters. This acts strongly in favour of prograde spin supply, see Fig \ref{fig:indcont3d} \textbf{C,D}.
    \item (Conversely) shorter encounter times for exterior pebbles (trajectory d2 to d4 and e1 to e3) ; and a smaller ``accretion height'' (lower $z$) due to the weakening gravity of the protoplanet with increasing vertical release distance.
\end{enumerate}

\subsection{Accreting larger pebbles and planetesimals}

\label{sec:DT}
Next, we consider varying the size of the accreted particles (in terms of the dimensionless stopping time $\tau_s$) while fixing the size of the body ($\zeta_w$).  We consider a range of $\tau_s \in [10^{-2},10^4]$ \added[]{in the case of zero inclination and eccentricity, $i = e = 0$}. The high $\tau_s$ limit effectively implies a vanishing gas-drag force and we expect therefore that our results must converge to the $i=0, e=0$ limit that have been previously calculated \citep{IdaNakazawa1990,DonesTremaine1993}.  Our results are presented in Figure 10, where, following \citet{DonesTremaine1993}, we have normalized with the product of the Hill velocity and the body radius $R$; $\left \langle \tilde{l_z} \right \rangle = \left \langle l_z \right \rangle / R_\mathrm{H} \Omega_0 R$ to compare results. 

For $0.01 \lesssim \tau_s \lesssim 0.1$ pebbles fall in symmetrically both from the interior and exterior bands since drift is negligible. However, the spin outcome is retrograde due to higher exterior flux because of shear. For $\tau_s \sim 0.1$ the co-rotation line eats away prograde accretion as discussed in previous sections. Combined with the higher exterior flux the spin outcome is even more retrograde. For $0.1 < \tau_s < 1$ back-accretion from the exterior band delivers high amounts of prograde spin due to increasing drift and higher exterior flux. For $\tau_s > 1$ drift decreases and chaotic bands are formed, meaning that the collision cross-section can be broken up by bands of missing trajectories \citep{IdaNakazawa1990}. Furthermore, the gas becomes less important and the co-rotation line therefore shifts more to the $x = 0$ line which supresses prograde contributions as pebbles follow parabolas from the interior band. The exterior pebbles are still moderately focused towards the planet by gas, leading to a strong retrograde infall. As both gas and radial drift effects diminish, the spin converges to the retrograde value found in previous works \citep{IdaNakazawa1990,LissauerSafronov1991,DonesTremaine1993}.

\section{Discussion}
\label{sec:Discussion}

\subsection{Caveats}
\label{sec:Global}
The results have been obtained in the approximation of a local shearing sheet patch for which Hill's approximations apply. This approach is perfectly valid for $x_0$ and $y_0$ $\ll$ than the orbital distance to the central body $r_0$ \citep{LiuOrmel2018}. The release distance $y_0$ increases for larger protoplanet radii due to the growing Hill radius to ensure that pebble trajectories are governed by gas-drag initially. If $y_0$ reaches a significant fraction of the orbital radius $r_0$, the curvature of the protoplanetary disk cannot be neglected anymore. 

To ensure the validity of our findings in the shearing sheet approximation, simulations are conducted in the global frame for the model $\alpha_p = 3 \times 10^{-3}$ and $\tau_s = 0.1$, see Figure \ref{fig:Glvsloc}. We applied the same procedure as \citet{LiuOrmel2018} who determined pebble accretion efficiency's in the global frame, referring to the global equations of motion of a pebble in the stellar frame and initial release conditions of the pebbles. The results in a global fashion are the same as our local simulations showing that the local frame produces valid outcomes for the intermediate to lower range of Stokes numbers. We do note that results might start to deviate for Stokes numbers higher than $\tau_s = 1$ due to the transition to gas-free accretion.

Collisions among pebbles within the Hill sphere have been ignored, which may result in the formation of circumplanetary pebble disks \added[]{\citep{SchlichtingSari2007}}
\deleted{ Johansen \& Lacerda (2010)}. This would in particular be important for small pebbles and large pebble fluxes; if bodies formed from $\tau_s=0.1$ particle over a protracted period, however, collisions are unimportant. The timescale on which pebbles with mass $m_p = (4/3) \pi s^3 \rho_{\bullet s}$ and radius $s$ collide is given by:
\begin{equation}
t_\mathrm{coll} = (n \sigma \Delta v)^{-1},
\end{equation}
with $n = \rho_p / m_p$ the number density of pebbles, $\sigma = 4 \pi s^2$ the mutual cross-section of pebble colliders and $\Delta v$ the typical velocity dispersion of the pebbles. Comparing this to the settling time of a pebble $t_\mathrm{sett} = b_\mathrm{coll} / v_\mathrm{sett}$ with $b_\mathrm{coll}$ the impact parameter, gives:
\begin{equation}
    N = \frac{t_\mathrm{sett}}{t_\mathrm{coll}} = \frac{3b_\mathrm{coll} \rho_p}{\rho_{\bullet s} s} \frac{\Delta v}{v_\mathrm{sett}},
\end{equation}
with $N$ the number of collisions pebbles undergo before settling onto the protoplanet. In terms of the disk parameters this can be written as:

\begin{equation}
\added[]{N \sim 0.05 \left (  \frac{q_p}{10^{-9}}\right  )^{1/3}\left ( \frac{\tau_s}{0.1} \right )^{2/3}\left ( \frac{h}{0.05}  \right )  \left (\frac{\rho_p}{\rho_g} \right )
 \left (\frac{\Delta v}{v_\mathrm{settl}} \right  )
},
\end{equation}
\added[]{where $q_p=M_p/M_\star$ and $h=H/r_0$ is the disk aspect ratio. Furthermore we assume that $v_\mathrm{th}\sim c_s = H\Omega_0$, pebbles are in the Epstein regime, and that the Hill limit applies: $b_\mathrm{coll} \sim R_\mathrm{H} \tau^{1/3}$ \citep{OrmelKlahr2010}. For asteroids $N \sim 0.05$, unless $\tau$ becomes very small. On the other hand, for larger bodies pebbles take increasingly longer to spiral in to the accreting body increasing $\rho_p/\rho_g$. This renders it likely that a disk will form around the body analogous to \citet{JohansenLacerda2010}. Thus, depending on the parameters collisions could become important and the effect on imparted spin should be considered in future work. }

\added[]{In this work pebble feedback on the gas has not been considered. Indeed, if the solid to gas ratio is of the order $\rho_p / \rho_g \gtrsim 1$ this alters dynamics of pebbles and the gas as the latter is forced to accrete with the solids in a prograde disk around the central body \citep{JohansenLacerda2010}. However, in the pebble accretion regime pebbles spiral in rapidly to the protoplanet with the radial velocity $v_r \sim 1/r^2$. This indicates that $\rho_p$ should not drastically increase because of the high accretion flux. Furthermore, our study is focused towards a more general description of spin delivery in the pebble accretion framework by considering the net sum of individual spin contributions of pebbles. We therefore have ignored collective effects for simplicity.} 

\added[]{We assume no initial rotation of the protoplanet in determining mean spin outcomes. However, bodies with radii $R \lesssim 100$ km experience a growth barrier as they are too inefficient to accrete pebbles \citep{VisserOrmel2016}. The angular momentum imparted during planetesimal formation could change the total spin outcomes obtained in the pebble accretion framework. Recent numerical results show however that trans-neptunian binaries forming through the streaming instability also preferentially orbit prograde around their mutual center of mass \citep{Nesvornyetal2019}. Furthermore we can assess the importance of the imprint of planetesimal spin by considering the total angular momentum supplied to a body of mass $M$. This quantity is given by $L = \int \left \langle l_z \right \rangle dm$. Rewriting this expression in logarithmic space gives $L = \int \left \langle l_z \right \rangle m \ d\log m$, showing that the integration of the mean angular momentum in Figure \ref{fig:figure2} is weighted by mass. Assuming that most of the mass of asteroids and protoplanets come from subsequent PA, rather than the initial planetesimal formation process, the spin state of these bodies will be given by Figure \ref{fig:figure2}.
} 

\added[]{When atmospheres form around the body, the gas flow becomes more complex \citep{Ormeletal2015A,Ormeletal2015B,Cimerman2017,LambrechtsLega2017, KurokawaTanigawa2018,Popovasetal2018}. These complicated but more realistic sub-structures of gas streamlines within the protoplanet Hill sphere are not considered in this work. Atmospheric structures do become important if $R_b/R \gtrsim 1$, where $R_b = GM / c_s^2$ is the Bondi radius of the body. For $R_b/R \gtrsim 3$ we estimate that the atmosphere has become dense enough to quantitatively alter the pebble trajectories and hence the spin contributions. While these effects are important to consider in future work for planets, the spin outcomes we obtain for asteroids are unaffected since $R_b/R \lesssim 1$.}
\section{Solar system bodies}
\label{sec:SSB}
For asteroids, we can use our general results presented in Figure \ref{fig:figure2} to constrain the conditions that operated during pebble accretion. First, we require relatively large pebbles $\tau_s \sim 0.1$; much smaller pebbles would not impart a significant net angular momentum, since they barely drift and shear across the impact cross section is small. Silicates particles, being less sticky, generally have a much lower $\tau_s$. \added[]{It is generally assumed that icy particles stick more readily and at larger collision velocities than silicate particles, so the need for relatively large pebbles could point us to icy pebbles.  However, a set of experiments by \citet{MusiolikWurm2019} indicate that the improved sticking only occurs in a relatively narrow temperature range at temperatures above $\sim 170 \ K$, so this enhancement might only be relevant near the snow line if these results turn up to hold in general. Also, a reduction of the disk headwind lowers relative velocities between grains possibly stimulating growth to larger (icy) pebbles. This would be consistent with the finding that (large) asteroids were icy bodies with a significant post-formation processing of the ice, making them dry \citep{SchmidtEtal2009}. ALMA observations also provide indications of the presence of decimeter sized pebbles in relatively young disks \citep{Zhangetal2015}. Nevertheless, the sticking properties of ice remain controversial and it is important to pursue further experimental and numerical research on the topic in the near future.} Second, for pebble accretion to operate on bodies $\sim$100 km we infer that the disk headwind parameter $v_\mathrm{hw}$ cannot be too high. Our standard value of $30\, \mathrm{m\,s}^{-1}$ already implies pebble accretion will fail for bodies smaller than $200$ km in radius (assuming $\tau_s=0.1$). However, the disk headwind is rather uncertain and could have been much smaller, in which case pebble accretion would operate on a much larger range of asteroids. This finding is again consistent with a cold formation environment.

For asteroids, the post-formation sublimation of ice could alter the rotation state of asteroids, although it is unlikely that this affects the rotation rate of large asteroids \citep{Jewitt1997}. For planets, our results are in good agreement with the observed spin periods of both the terrestrial planets Earth and Mars, and the ice giants Uranus and Neptune. Not only does this suggest that pebble accretion naturally provides systematic spins to (exo)solar system bodies, it provides an additional validation for pebble accretion as a planet formation mechanism. Undoubtedly, impacts, outgassing, long-term chaotic evolution, and the formation of a planet atmosphere, which leads to different pebble aerodynamics and eventually ablation of pebbles\citep{BrouwersEtal2018} have shaped the periods and orientation of the spin axes of the solar system's planetary bodies.  Nonetheless, the evidence in the solar system renders it likely that bodies started off with a systematic vertical spin provided by pebble accretion. 
\section{Conclusions}
\label{sec:concl}
We have calculated the spin angular momentum per unit mass that pebbles supply to a protoplanet during accretion, both in a 2D and 3D fashion. The net angular momentum gained by the protoplanet is obtained by summing individual pebble supply to the body at impact. The main conclusions of the obtained results are as follows:

\begin{enumerate}
    \item Pebble accretion can deliver significant systematic rotation to an accreting body for a large part of the parameter space. In our results, spin values reach the value of spin periods typically observed in the Solar System both in magnitude and direction. The results are in good agreement with the observed spin rotation of the larger asteroids such as for example Ceres, Pallas, Vesta, and Hygiea. 
    \item The cause of the asymmetry in spin accumulation by the accreting body, is a discrepancy in encounter times of pebbles that accrete interior and pebbles that accrete exterior to the collision cross-section. The difference in encounter times follows from the asymmetric infall angle of pebbles due to radial drift around the onset of pebble accretion and the large discrepancy over the collision cross-section due to keplerian shear. The asymmetry favors prograde spin rotation by the accreting body and, in a smaller region of the parameter space, favors retrogade spin rotation. 
     \item The degree of asymmetry between prograde and retrograde spin contributions, $\delta = \langle l_z \rangle / \sqrt{\langle l_z^2 \rangle}$ reaches values of $\delta \sim \pm 0.3$. This is significantly higher than $\delta = 10^{-3}$ that has been found in gas-free planetesimal accretion.
    \item In 3D there is a propensity for prograde rotation, because the pebbles are accreted preferentially from the (prograde) side characterized by long lasting encounters.
    
    \item In generally, small pebbles ($\tau_s \ll 0.1$) bestow much smaller spin than larger pebbles, due to the more symmetric impact geometry. Applying our results to asteroids, we infer that they formed in a cold environment and accreted pebbles with significant ice fractions. Much of the ice was subsequently lost by post-formation processes.

\end{enumerate}
\bibliographystyle{model5-names}
\bibliography{mybibfile}
\subsubsection*{Acknowledgements}
This work has been financially supported by the Netherlands Organisation
for Scientific Research (NWO project no. ALWGO/15-01 and VIDI project 639.042.422)
\begin{table}[h!]
    \small
    \centering
    \caption{\textbf{Relation between the dimensionfull (Dim.full) and dimensionless (Dim.less) quantities}}
\begin{tabular}{lllll}
\hline
\hline
Quantity                & Dim.full        & Dim.less                               & Dim.full range               & Dim.less range                                          \\ \hline
Hill radius             & $R_\mathrm{H}$  & 1                                       &                                &                                                              \\
Keplerian frequency     & $\Omega_0$      & 1                                       &                                &                                                              \\
Gas headwind velocity   & $v_\mathrm{hw}$ & $\zeta_w = v_\mathrm{hw}/R_H\Omega_0$ & $30 \ \mathrm{ms^{-1}}$        & 50--0.5                                                     \\
Protoplanet radius      & $R$ & $\alpha_p = R/R_\mathrm{H}$             & $10 \ \mathrm{km} - 7000 \ \mathrm{km}$ & $2,3,7.5 (\times 10^{-3})$ \\
Stopping time           & $t_s$           & $\tau_s = t_s\Omega_0$      &                                & 0.01--0.5                                                   \\
Specific mean frequency & $\omega_z$      & $\omega_z / \omega_\mathrm{crit}$       &                                &                                                              \\ \hline

\end{tabular}
\label{tab:tabledimvsdimless}
\end{table}

\section*{Appendix A.}
\label{sec: appA}
\subsection*{Asteroid Lightcurve Data}
Diameter, latitude and period are obtained from the Asteroid Lightcurve Data Base (LCDB)
V2.0\citep{WarnerEtal2009}. The density is taken from \citet{BaerEtal2012}. If unknown $\rho$ = 2 g $\mathrm{cm}^{-3}$ is used.
\begin{table}[h!]
\centering

\begin{tabular}{lrrrrrr}
Asteroid              & (equivalent)          & density               & ecliptic latitude     & period                & $\omega/\omega_\mathrm{crit}$& \\
                      & diameter [km]         & [g/cc]                & polar axis $\beta$    & [h]                   &                       & \\
\hline 
Ceres                 & {$970$}               & {$2.1$}               & {$82$}                & {$9.1$}               & {$0.25$}              & \\
Pallas                & {$510$}               & {$2.6$}               & {$-13$}               & {$7.8$}               & {$0.26$}              & \\
Juno                  & {$250$}               & {$2.7$}               & {$20$}                & {$7.2$}               & {$0.28$}              & \\
Vesta                 & {$470$}               & {$3.4$}               & {$42$}                & {$5.3$}               & {$0.33$}              & \\
Hebe                  & {$190$}               & {$4.0$}               & {$50$}                & {$7.3$}               & {$0.23$}              & \\
Iris                  & {$200$}               & {$2.2$}               & {$19$}                & {$7.1$}               & {$0.31$}              & \\
Metis                 & {$200$}               & {$2.4$}               & {$20$}                & {$5.1$}               & {$0.42$}              & \\
Hygiea                & {$350$}               & {$2.1$}               & {$-35$}               & {$28$}                & {$0.082$}             & \\
Parthenope            & {$160$}               & {$3.3$}               & {$16$}                & {$14$}                & {$0.13$}              & \\
Egeria                & {$210$}               & {$3.4$}               & {$20$}                & {$7.0$}               & {$0.25$}              & \\
Irene                 & {$150$}               & {$2.0$}               & {$-14$}               & {$15$}                & {$0.15$}              & \\
Eunomia               & {$260$}               & {$2.8$}               & {$-67$}               & {$6.1$}               & {$0.32$}              & \\
Psyche                & {$220$}               & {$7.3$}               & {$-8.0$}              & {$4.2$}               & {$0.29$}              & \\
Fortuna               & {$220$}               & {$1.4$}               & {$69$}                & {$7.4$}               & {$0.38$}              & \\
Themis                & {$200$}               & {$2.8$}               & {$69$}                & {$8.4$}               & {$0.24$}              & \\
Amphitrite            & {$210$}               & {$3.0$}               & {$-26$}               & {$5.4$}               & {$0.36$}              & \\
Euphrosyne            & {$280$}               & {$6.6$}               & {$66$}                & {$5.5$}               & {$0.23$}              & \\
Laetitia              & {$160$}               & {$3.2$}               & {$32$}                & {$5.1$}               & {$0.36$}              & \\
Daphne                & {$190$}               & {$2.4$}               & {$-32$}               & {$6.0$}               & {$0.36$}              & \\
Eugenia               & {$210$}               & {$1.1$}               & {$-36$}               & {$5.7$}               & {$0.55$}              & \\
Doris                 & {$220$}               & {$2.1$}               & {$57$}                & {$12$}                & {$0.19$}              & \\
\end{tabular}
\end{table}
\begin{table}
\centering
\begin{tabular}{lrrrrrr}
Asteroid              & (equivalent)          & density               & ecliptic latitude     & period                & $\omega/\omega_\mathrm{crit}$& \\
                      & diameter [km]         & [g/cc]                & polar axis $\beta$    & [h]                   &                       & \\
\hline 
Europa                & {$290$}               & {$1.6$}               & {$36$}                & {$5.6$}               & {$0.47$}              & \\
Cybele                & {$240$}               & {$0.99$}              & {$-3.0$}              & {$6.1$}               & {$0.55$}              & \\
Freia                 & {$160$}               &                       & {$12$}                & {$10$}                & {$0.19$}              & \\
Io                    & {$160$}               &                       & {$-68$}               & {$6.9$}               & {$0.28$}              & \\
Sylvia                & {$260$}               & {$1.2$}               & {$67$}                & {$5.2$}               & {$0.58$}              & \\
Thisbe                & {$200$}               & {$3.1$}               & {$68$}                & {$6.0$}               & {$0.31$}              & \\
Minerva               & {$150$}               &                       & {$21$}                & {$6.0$}               & {$0.32$}              & \\
Aurora                & {$170$}               &                       & {$3.0$}               & {$7.2$}               & {$0.26$}              & \\
Camilla               & {$220$}               & {$1.4$}               & {$55$}                & {$4.8$}               & {$0.58$}              & \\
Lachesis              & {$170$}               &                       & {$39$}                & {$47$}                & {$0.041$}             & \\
Hermione              & {$210$}               & {$1.4$}               & {$13$}                & {$5.6$}               & {$0.51$}              & \\
Elektra               & {$200$}               & {$1.3$}               & {$-88$}               & {$5.2$}               & {$0.55$}              & \\
Juewa                 & {$160$}               &                       & {$50$}                & {$21$}                & {$0.091$}             & \\
Nuwa                  & {$150$}               &                       & {$25$}                & {$8.1$}               & {$0.23$}              & \\
Hilda                 & {$170$}               &                       & {$29$}                & {$6.0$}               & {$0.32$}              & \\
Bertha                & {$190$}               &                       & {$34$}                & {$25$}                & {$0.076$}             & \\
Loreley               & {$170$}               &                       & {$31$}                & {$7.2$}               & {$0.26$}              & \\
Ino                   & {$150$}               &                       & {$-14$}               & {$6.2$}               & {$0.31$}              & \\
Eunike                & {$160$}               &                       & {$4.0$}               & {$22$}                & {$0.087$}             & \\
Ismene                & {$160$}               &                       & {$23$}                & {$6.5$}               & {$0.29$}              & \\
Dido                  & {$160$}               &                       & {$37$}                & {$5.7$}               & {$0.33$}              & \\
Germania              & {$170$}               &                       & {$44$}                & {$16$}                & {$0.12$}              & \\
\end{tabular}
\end{table}
\begin{table}
\centering
\begin{tabular}{lrrrrrr}
Asteroid              & (equivalent)          & density               & ecliptic latitude     & period                & $\omega/\omega_\mathrm{crit}$& \\
                      & diameter [km]         & [g/cc]                & polar axis $\beta$    & [h]                   &                       & \\
\hline 
Bamberga              & {$230$}               & {$1.7$}               & {$68$}                & {$29$}                & {$0.086$}             & \\
Chicago               & {$160$}               &                       & {$-59$}               & {$7.4$}               & {$0.26$}              & \\
Eleonora              & {$160$}               &                       & {$41$}                & {$4.3$}               & {$0.45$}              & \\
Palma                 & {$190$}               &                       & {$-5.0$}              & {$8.6$}               & {$0.22$}              & \\
Siegena               & {$170$}               &                       & {$26$}                & {$9.8$}               & {$0.20$}              & \\
Aspasia               & {$170$}               &                       & {$30$}                & {$9.0$}               & {$0.21$}              & \\
Diotima               & {$180$}               &                       & {$4.0$}               & {$4.8$}               & {$0.40$}              & \\
Patientia             & {$230$}               & {$3.4$}               & {$21$}                & {$9.7$}               & {$0.18$}              & \\
Davida                & {$300$}               & {$3.0$}               & {$24$}                & {$5.1$}               & {$0.37$}              & \\
Herculina             & {$220$}               & {$5.8$}               & {$11$}                & {$9.4$}               & {$0.15$}              & \\
Hektor                & {$230$}               &                       & {$-32$}               & {$6.9$}               & {$0.28$}              & \\
Interamnia            & {$320$}               & {$2.3$}               & {$66$}                & {$8.7$}               & {$0.25$}              & \\
Winchester            & {$170$}               &                       & {$-60$}               & {$9.4$}               & {$0.20$}              & \\
Berbericia            & {$150$}               &                       & {$25$}                & {$7.7$}               & {$0.25$}              & \\
Hispania              & {$160$}               & {$1.7$}               & {$49$}                & {$15$}                & {$0.17$}              & \\
Pholus                & {$170$}               &                       & {$30$}                & {$10$}                & {$0.19$}              & \\
Bienor                & {$160$}               &                       & {$50$}                & {$9.1$}               & {$0.21$}              & \\
\hline 
\end{tabular}
\end{table}

\begin{figure*}[t]
\centering
\includegraphics[width=1\textwidth]{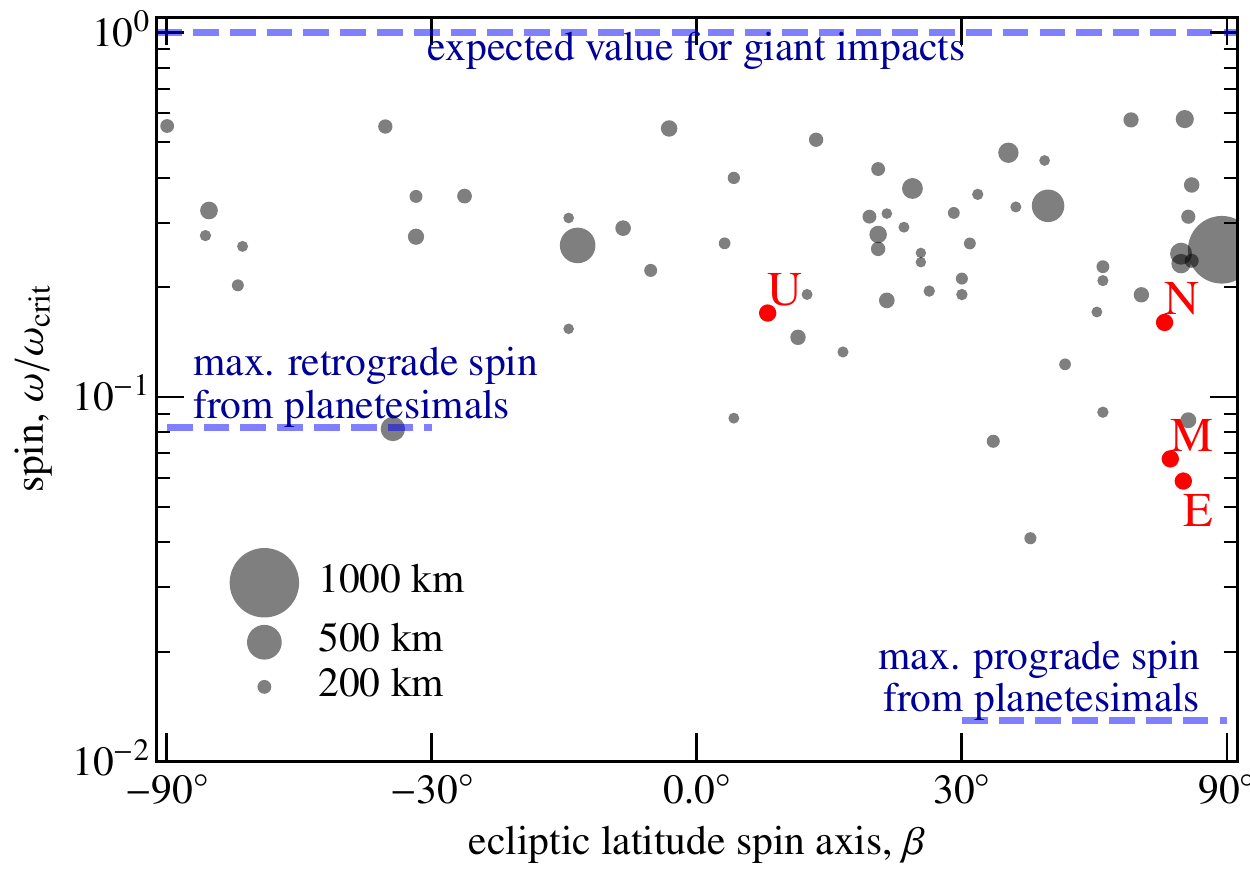}
\caption{\label{fig:figure1} \textbf{Spin rate and orientation of planets (Earth, Mars, Uranus, and Neptune indicated with E, M, U, N, respectively) and asteroids of diameter D larger than 150 km (circles)}. The spin frequency in terms of the critical breakup frequency $\omega_\mathrm{crit}=\sqrt{GM_p/R^3}$ is plotted against the orientation of the spin vector to the ecliptic, where $\beta$ measures ecliptic latitude. 
%For the planets, Earth's angular momentum is that of the Earth-Moon system (\ccc{TBD (chris!)}); Mercury and Venus have a negligible\? spin. 
For asteroids, the symbol size scales with the diameter. The giant impact scenario predicts a spin magnitude close to breakup and an isotropic orientation of spin vectors\citep{KokuboIda2007} ($\sin\beta$, which is proportional to the x-axis scaling, would then be uniformly distributed). The lower horizontal lines indicate the maximum retrograde and prograde spin rates obtained from systematic planetesimal accretion\citep{DonesTremaine1993}. Planetary spins are too large for planetesimal accretion and appear too much skewed towards $\beta=90^\circ$ for the giant impact scenario. See Appendix A for the asteroid properties.
}
\end{figure*}
\begin{figure*}[t]
    \centering
    \includegraphics[width=0.7\textwidth]{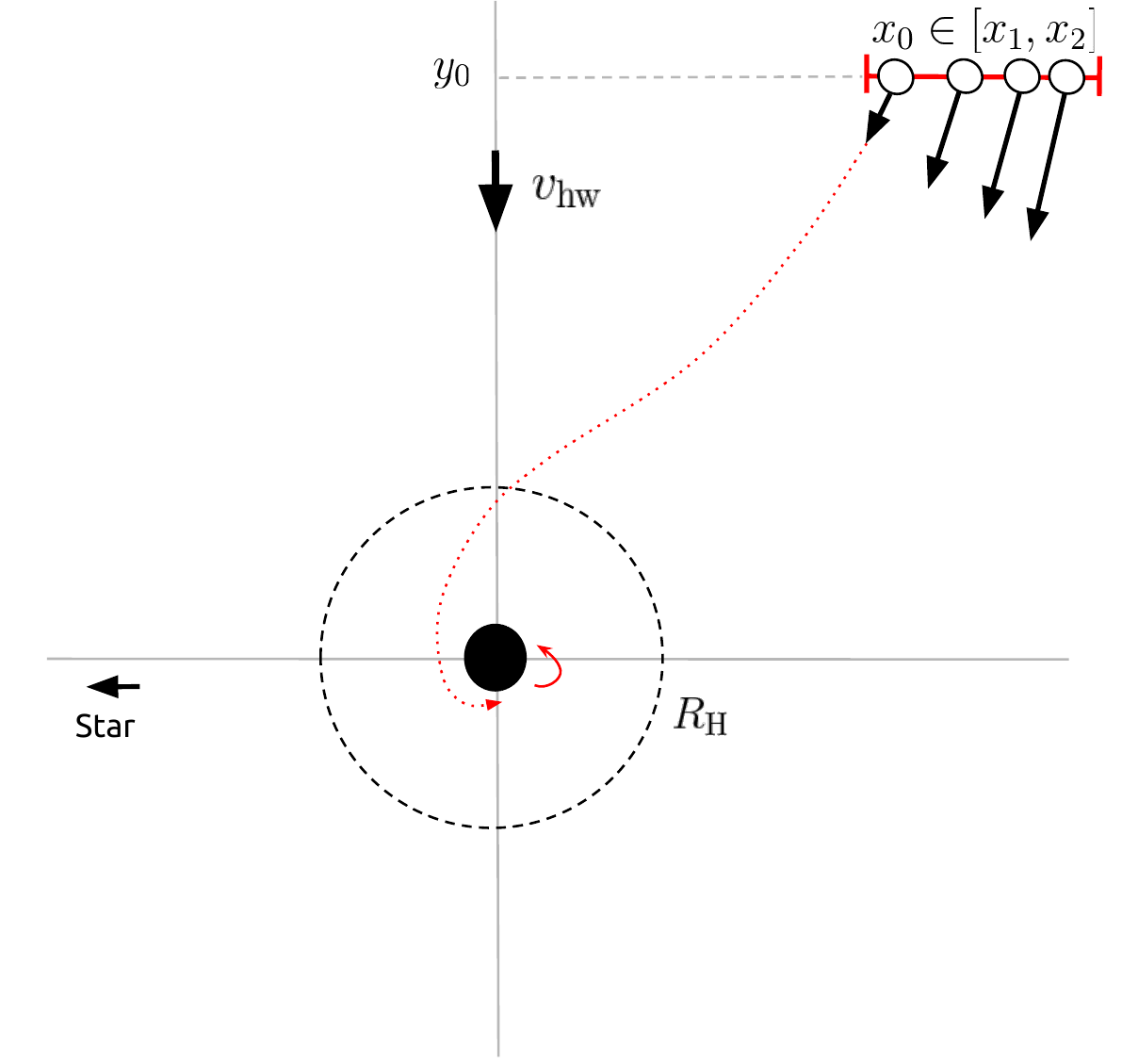}
    \caption{\textbf{A sketch of the shearing sheet domain co-rotating with a protoplanet (center).} Pebbles enter from the top at the right side from $(x_0,y_0)$ due to a negative shear velocity and the downwards pointing headwind. The flux of pebbles increases to the right due to the increasing shear velocity. The highlighted pebble (red dotted) supplies its angular momentum to the protoplanet at impact.}
    \label{fig:sketch}
\end{figure*}
\begin{figure}[t]
    \centering
    \includegraphics[width=10cm]{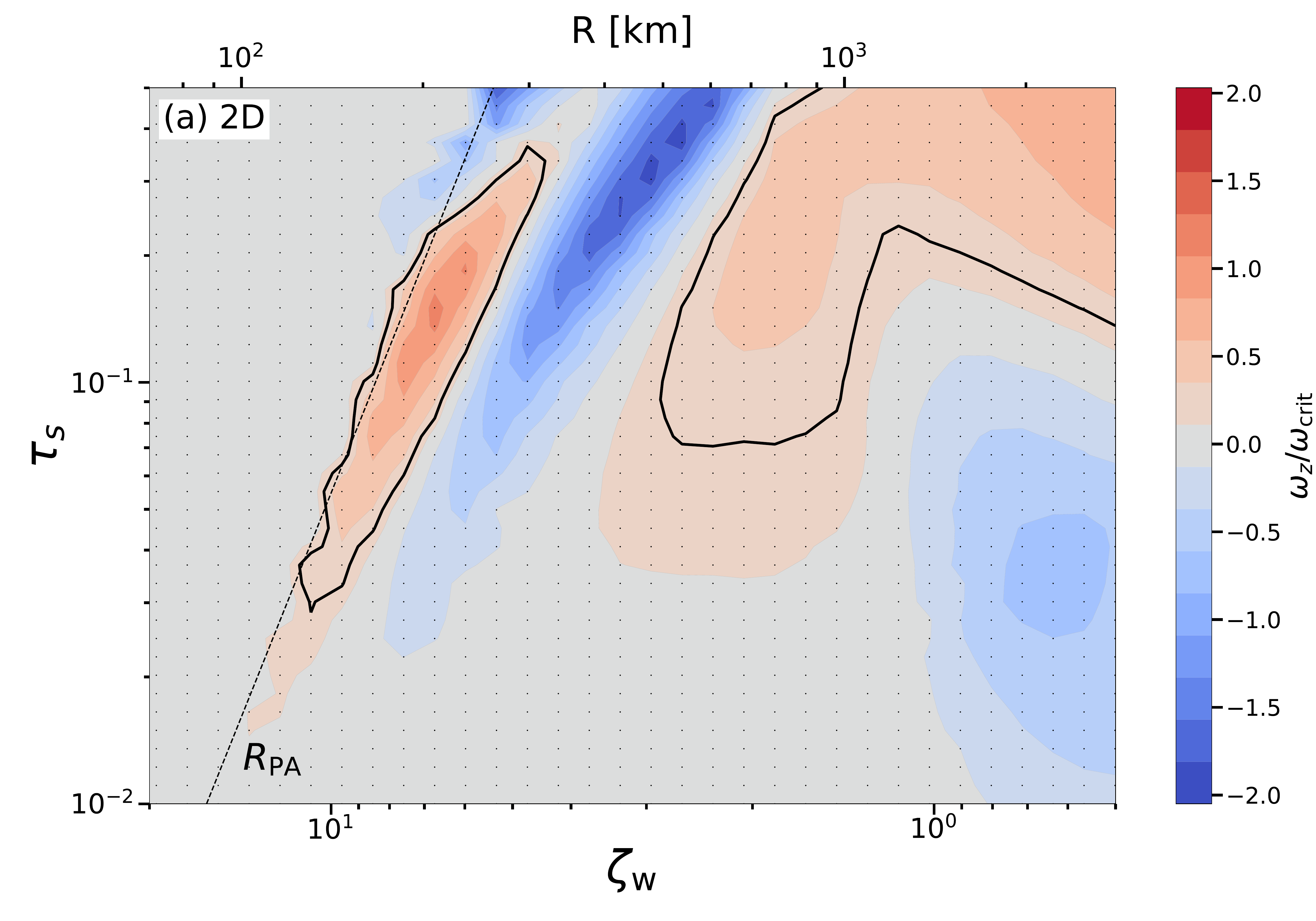}
    \includegraphics[width=10cm]{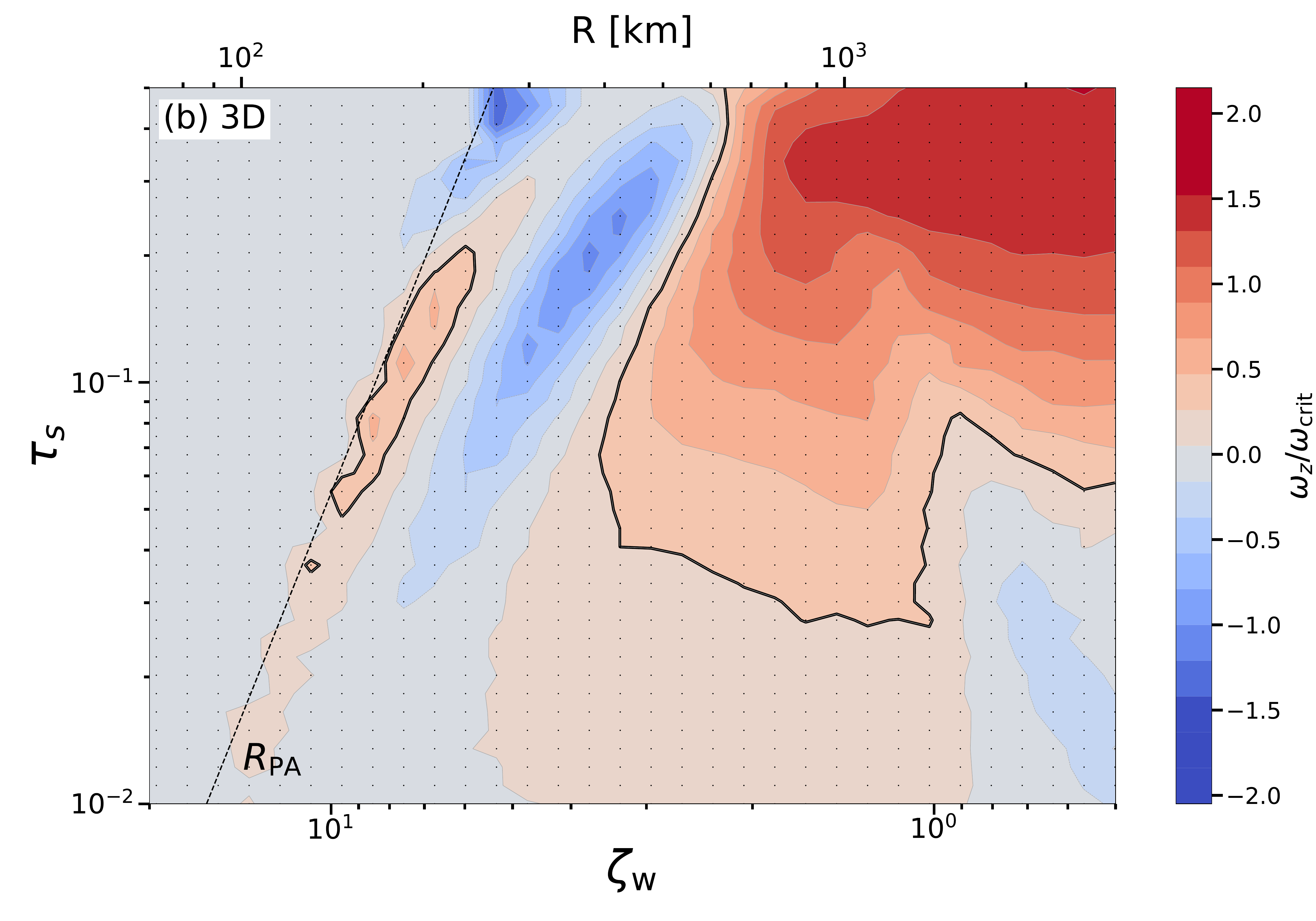}
    \caption{\label{fig:figure2} 
    \small
    \textbf{Spin rotation rates obtained from pebble accretion for the planar case (top) and 3D case (bottom)}. For each combination of stopping time $\tau_s$ and headwind-to-shear parameter $\zeta_w$ the average vertical angular momentum per unit mass $\langle l_z \rangle$ transfered by pebbles to the protoplanet is calculated. This is converted into a normalized spin through the relation $\omega/\omega_\mathrm{crit} = \langle l_z \rangle /n\sqrt{GM R}$ where an inertia factor of $n=0.4$ has been used. Valid for $\alpha_p=0.002$. The transformation from $\zeta_w$ to $R$ assumes a disk headwind of $v_\mathrm{hw} = 30$ m/s.
}
\end{figure}
\begin{figure}
    \includegraphics[width=6cm]{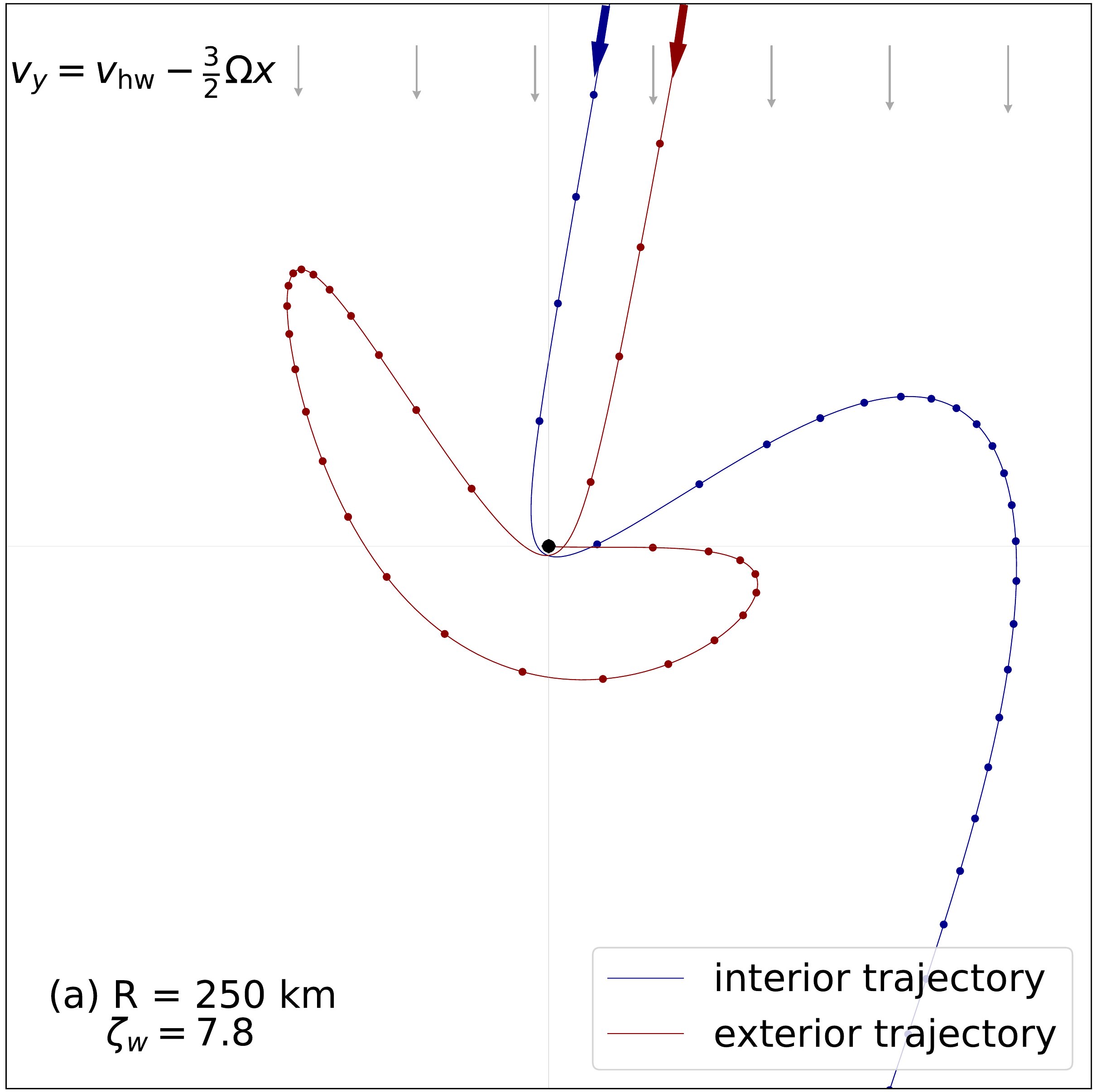}
    \includegraphics[width=6cm]{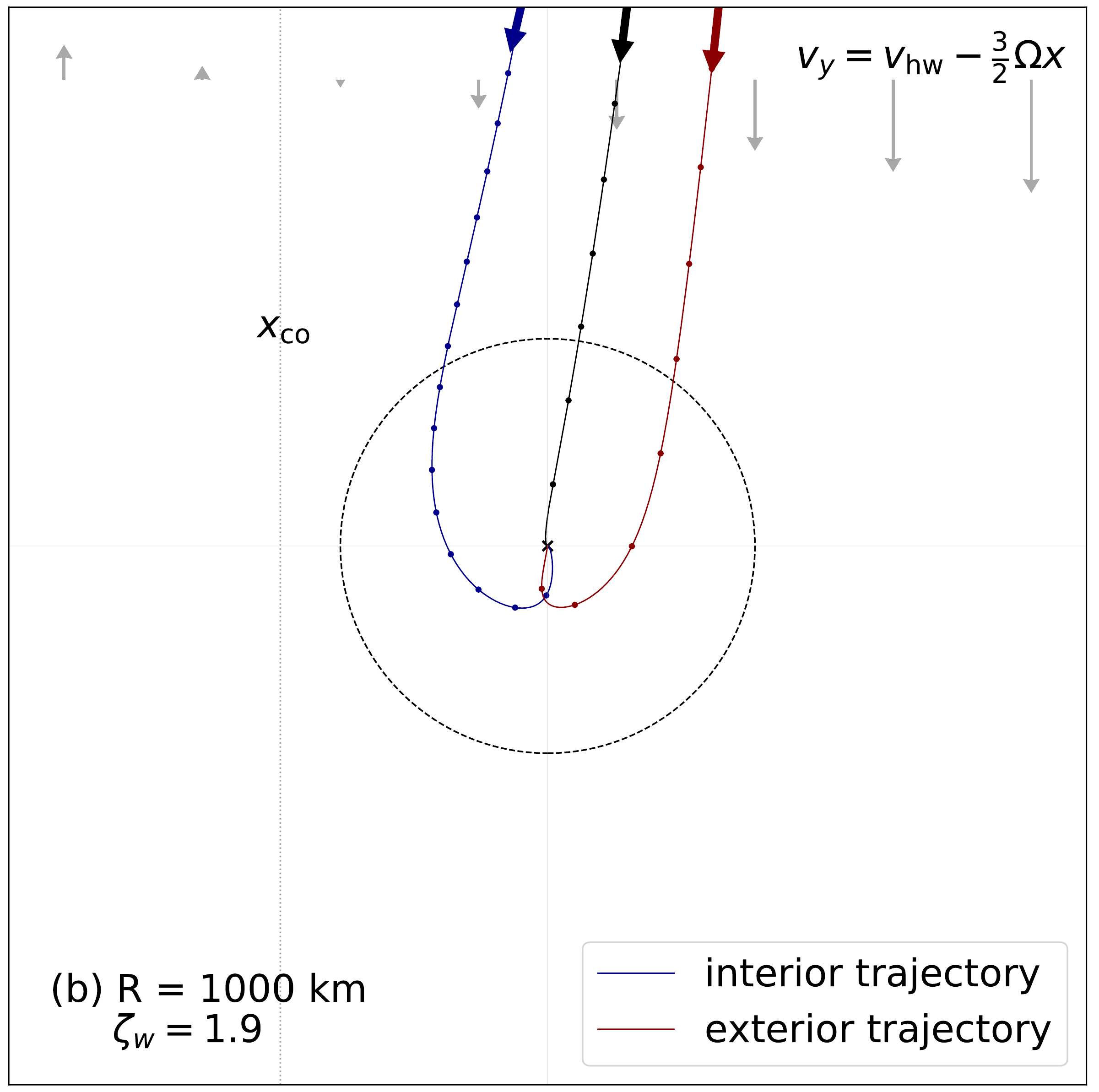}
    \caption{\label{fig:figure3} 
        \textbf{Preference of spin asymmetry explained by geometry of pebble accretion}. Pebble accreting trajectories (for $\tau_s=0.1$ and $\alpha_p = 3 \times 10^{-3}$ are plotted in the frame of the circularly-moving planet (center). In this frame pebbles approach the planet at an angle, due to their radially drift motion. The distance between two adjacent dots on the trajectory indicate the same amount of time.  (a) Due to the slanted approach pebbles encountering the planet (black dot) on an exterior orbit (red) will be scattered at a steeper angle with respect to the headwind than those pebbles that approach on the interior side (blue). Consequently, these pebbles face a stronger aerodynamic deceleration and the prolonged encounter duration allows their gravitational capture, whereas the more horizontally-deflected pebbles (blue) escape. In (b) the situation for an accretion range of the order of the planet Hill radius $R_\mathrm{H}$ (dashed circle) is illustrated, where the red and blue curves indicate the extreme trajectories that lead to capture. As the shear (arrows) is significant for larger planets, pebbles on interior (blue) trajectories approach at a lower velocity, resulting in longer encounters and a higher total capture rate by the planet. In both cases the asymmetry results in a significant net prograde contribution.
}
\end{figure}
\begin{figure*}[h!]
    \centering
    \includegraphics[width=1.1\textwidth]{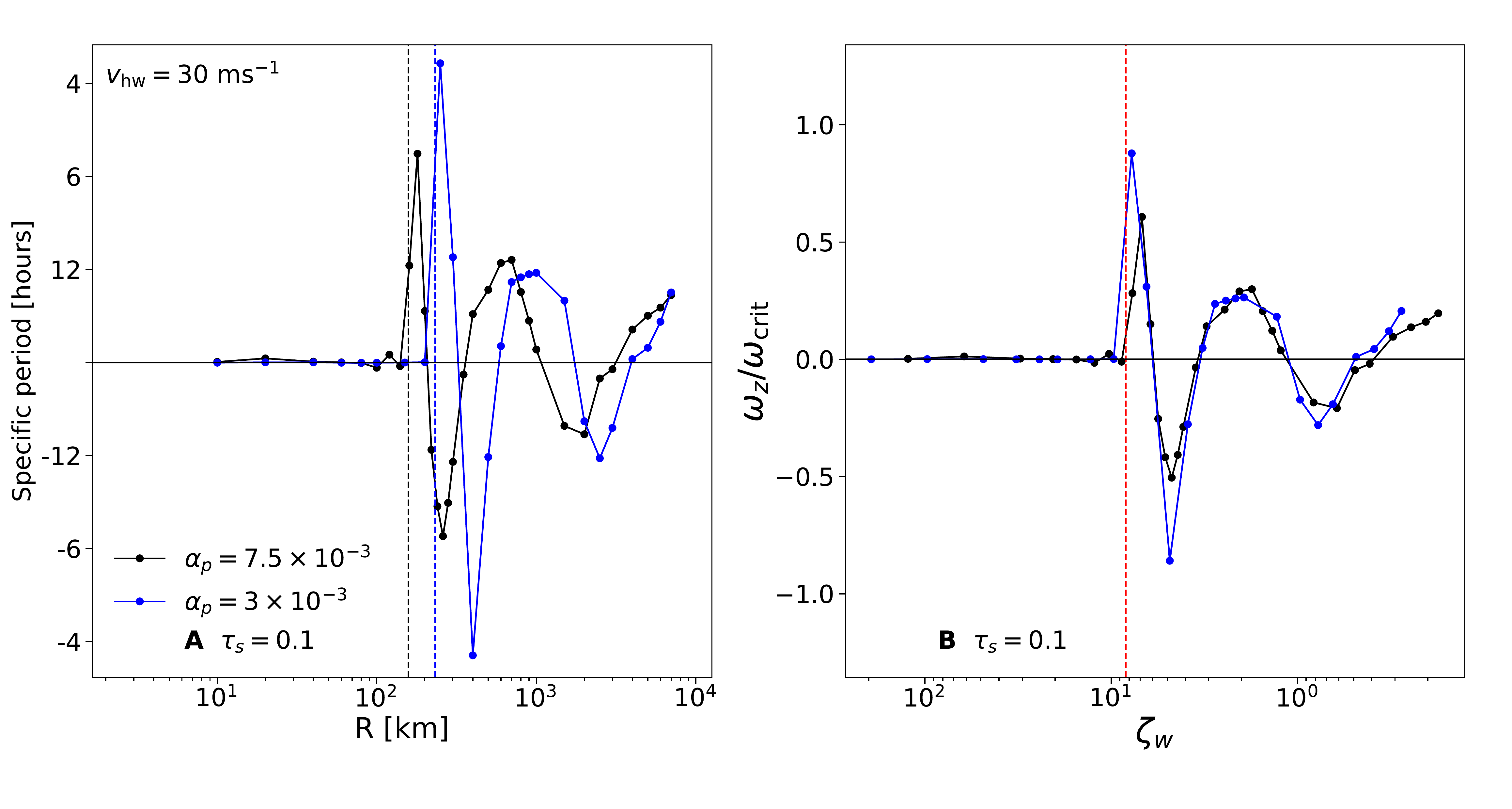}
    \caption{\textbf{ Spin results obtained in dimensionfull and dimensionless units}. \textbf{A}: The specific period for $\tau_s = 0.1$ for planetesimal/protoplanet sizes $R \in [10 \ \mathrm{km},7000 \ \mathrm{km}]$ and a protoplanet density of 1 g $\mathrm{cm^{-3}}$. The dashed vertical line shows the protoplanet radius where pebble accretion starts (Equation \ref{eq:onsetpa}). \textbf{B}:  the same data in terms of dimensionless quantities $\zeta_w$ vs the fraction of the break-up frequency. \deleted[]{The general shape of the curves is the same and overlap in \textbf{B}  but the amplitude varies with $\alpha_p$ and Stokes number $\tau_s$.}}
    \label{fig:lzgraph}
\end{figure*}

\begin{figure*}[p]
    \centering
    \includegraphics[width=1\textwidth]{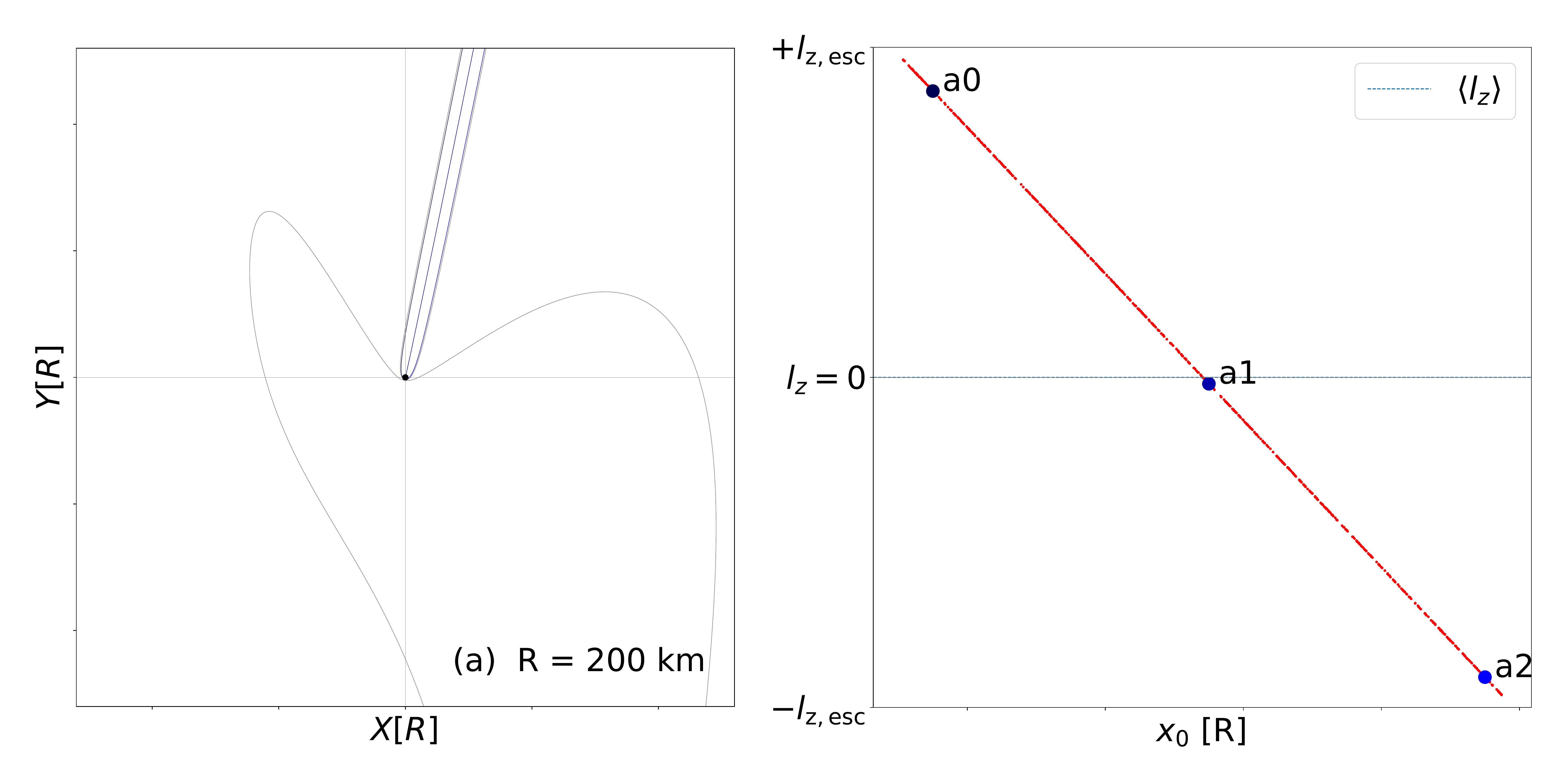}
    \includegraphics[width=1\textwidth]{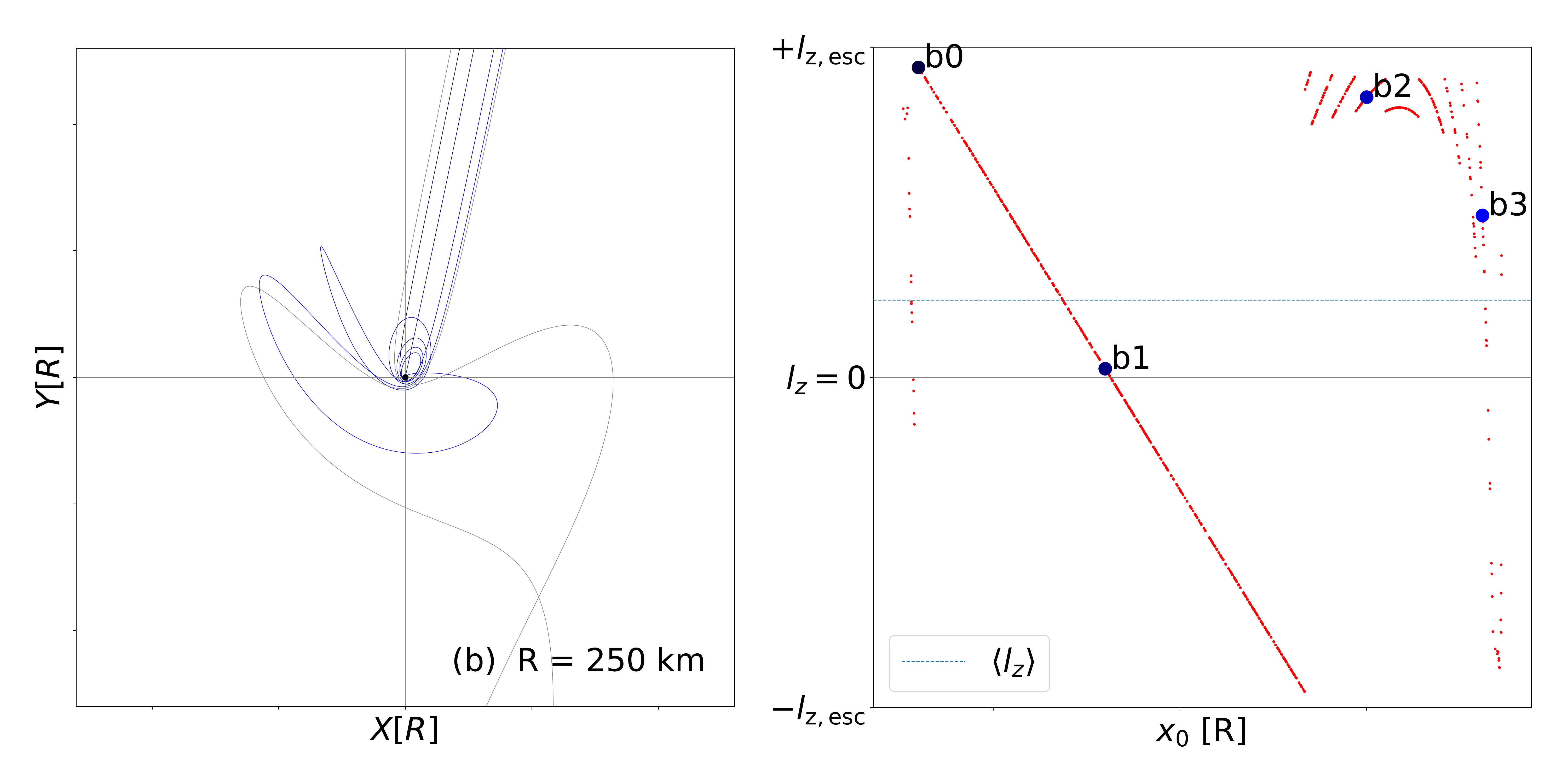}
    \includegraphics[width=1\textwidth]{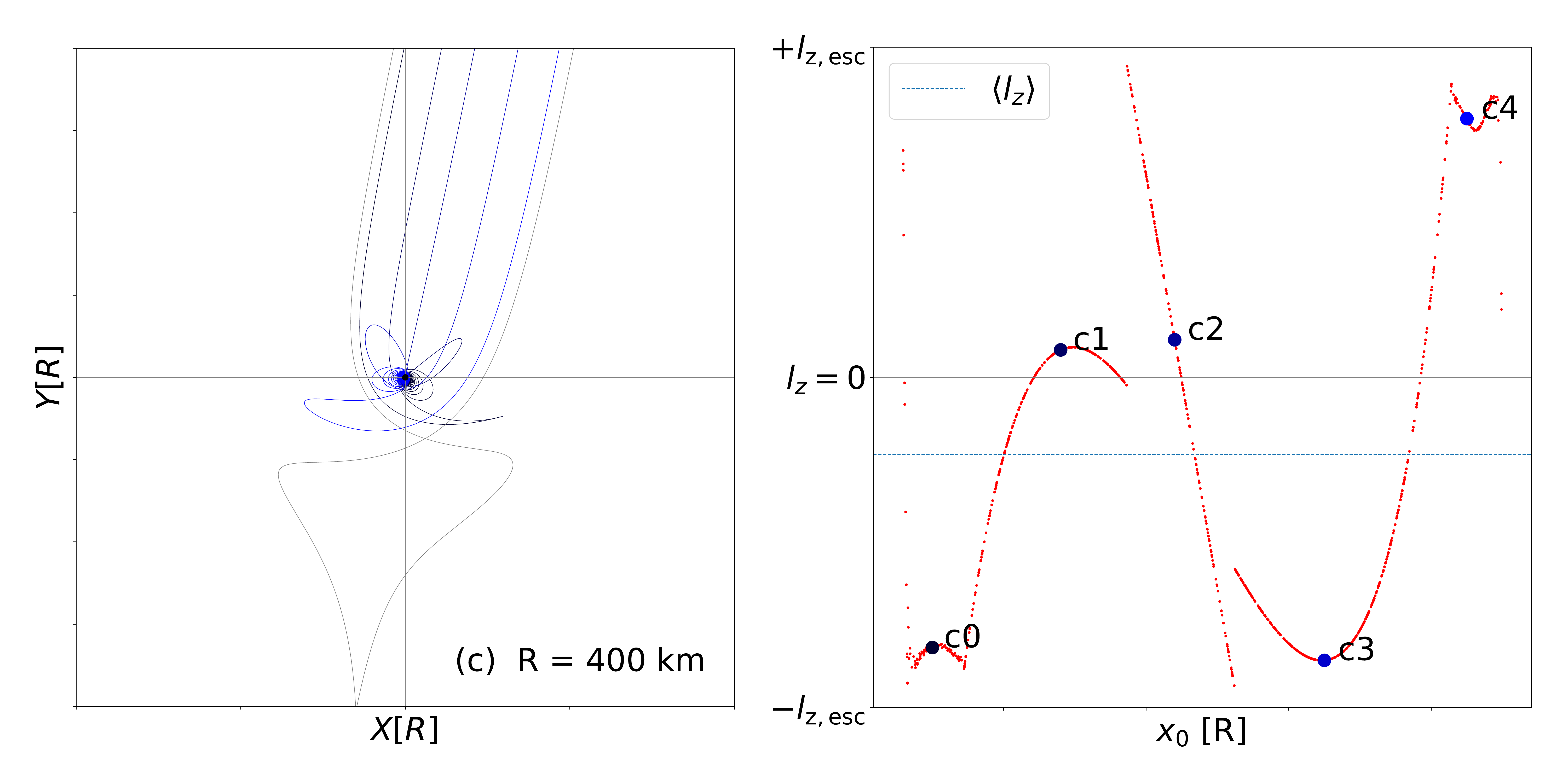}
    \label{fig:ballpebb}
\end{figure*}
\begin{figure*}[p]
    \centering
    
    \includegraphics[width=1\textwidth]{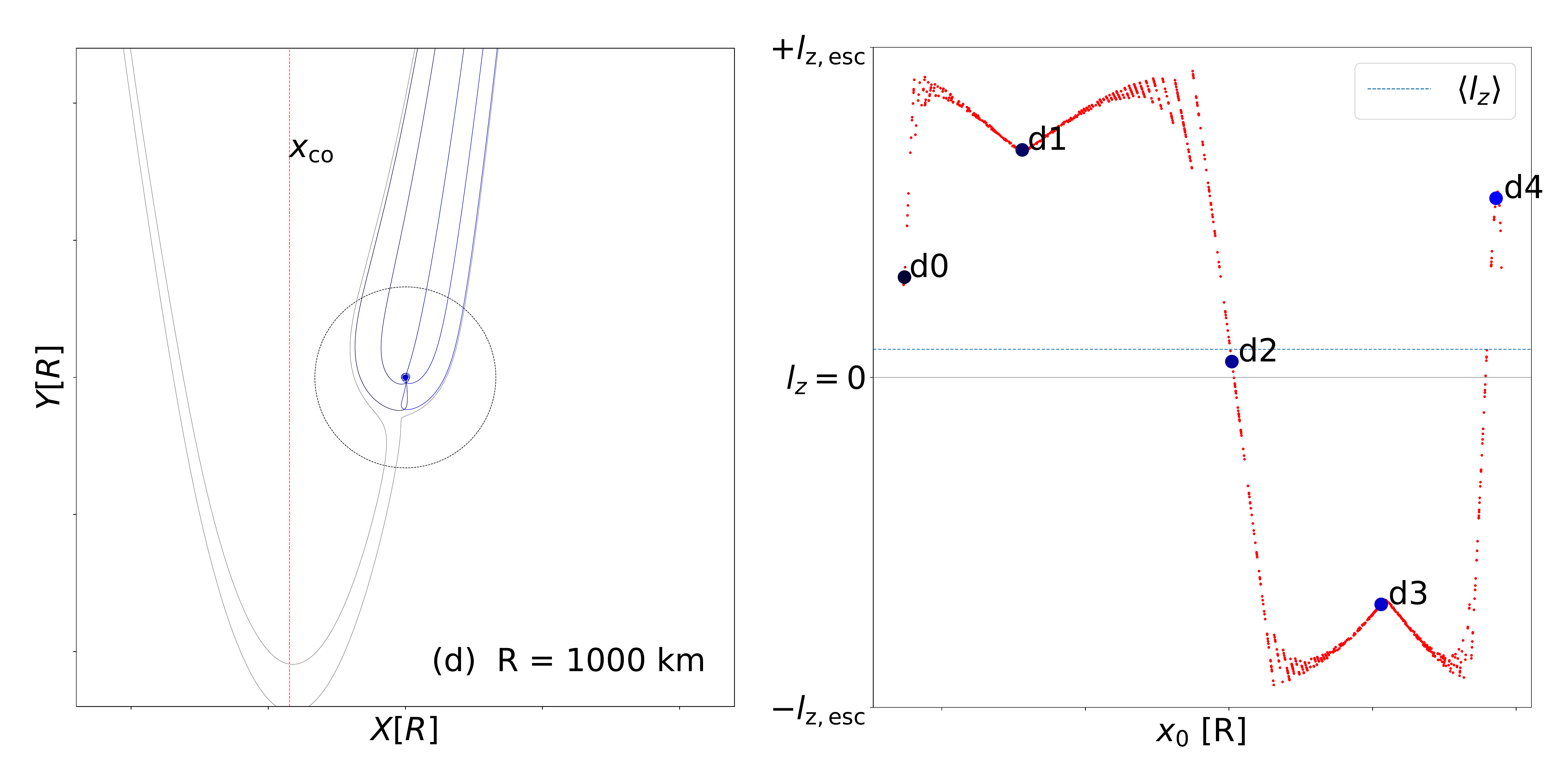}
    \includegraphics[width=1\textwidth]{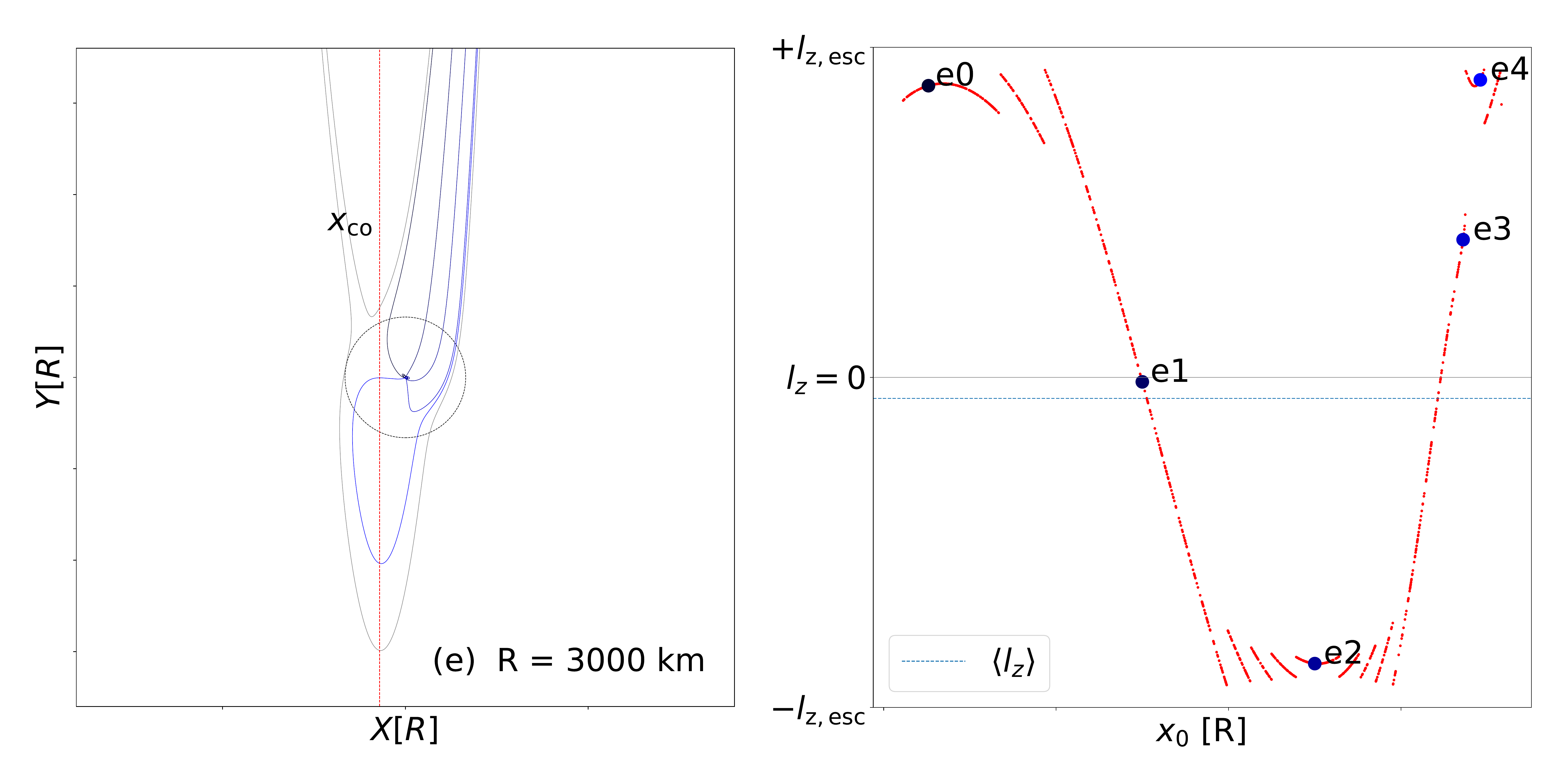}
    \caption{\small \textbf{Pebble trajectories (left) and corresponding spin curves (right)}. They are selected from the $\alpha_p = 3 \times 10^{-3}$ integrations and $\tau_s=0.1$ for indicated protoplanet radii. The spin curves show individual SAM contributions to the protoplanet versus initial release distance $x_0$. $x_0$ ranges from the interior side ($x_0=x_1$) to the exterior side ($x_0=x_2$) of the collision cross-section. The resulting mean spin $\left \langle {l_z} \right \rangle$ is indicated with the dashed blue line. The extent of the y-axis ranges from $\pm$ $l_\mathrm{z,esc}$. Shown pebble trajectories consist of two misses (gray trajectories) and a range of impacts. The impacting trajectories correspond from black (interior) to blue (exterior) with the indicated dots in the corresponding spin curve to clarify individual spin supply behavior. Additionally these selected trajectories are uniquely labeled for reference in the main text. For \textbf{D}, and \textbf{E}, the Hill sphere is indicated by the dashed circle and the co-rotation line with the red vertical line.}
    \label{fig:ballpebb}
\end{figure*}

\begin{figure*}[t]
    \includegraphics[width=\textwidth]{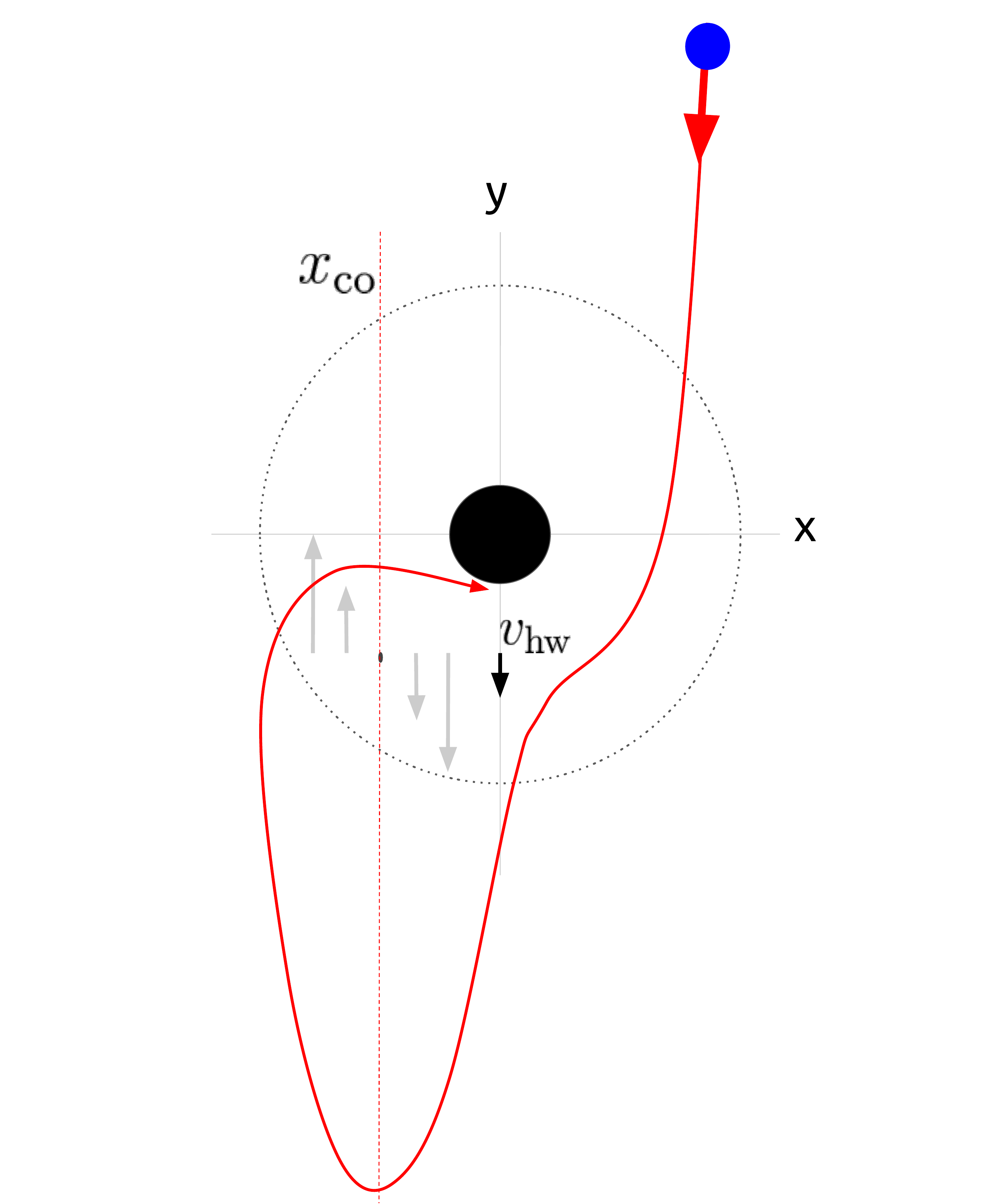} 
    \caption{\textbf{Clarifying illustration of a secondary Hill sphere encounter of a pebble.} At first approach the pebble escapes from the Hill sphere and drifts to the co-rotation line at which the shear + headwind (gray arrows) are zero. The headwind velocity is indicated by the black arrow. As the pebble approaches the Hill sphere for a second time, the pebble is pulled inwards by the protoplanet gravity. During the path from $-R_\mathrm{Hill}$ to the protoplanet surface, the pebble is simultaneously accelerated downwards due to the negative shear velocity and headwind velocity. As a result the pebble is forced to impact counterclockwise below the protoplanet delivering prograde spin rotation.}
    \label{fig:Horseshoe}
\end{figure*}
\begin{figure*}[t]
    \centering
    \includegraphics[width=1.2\textwidth]{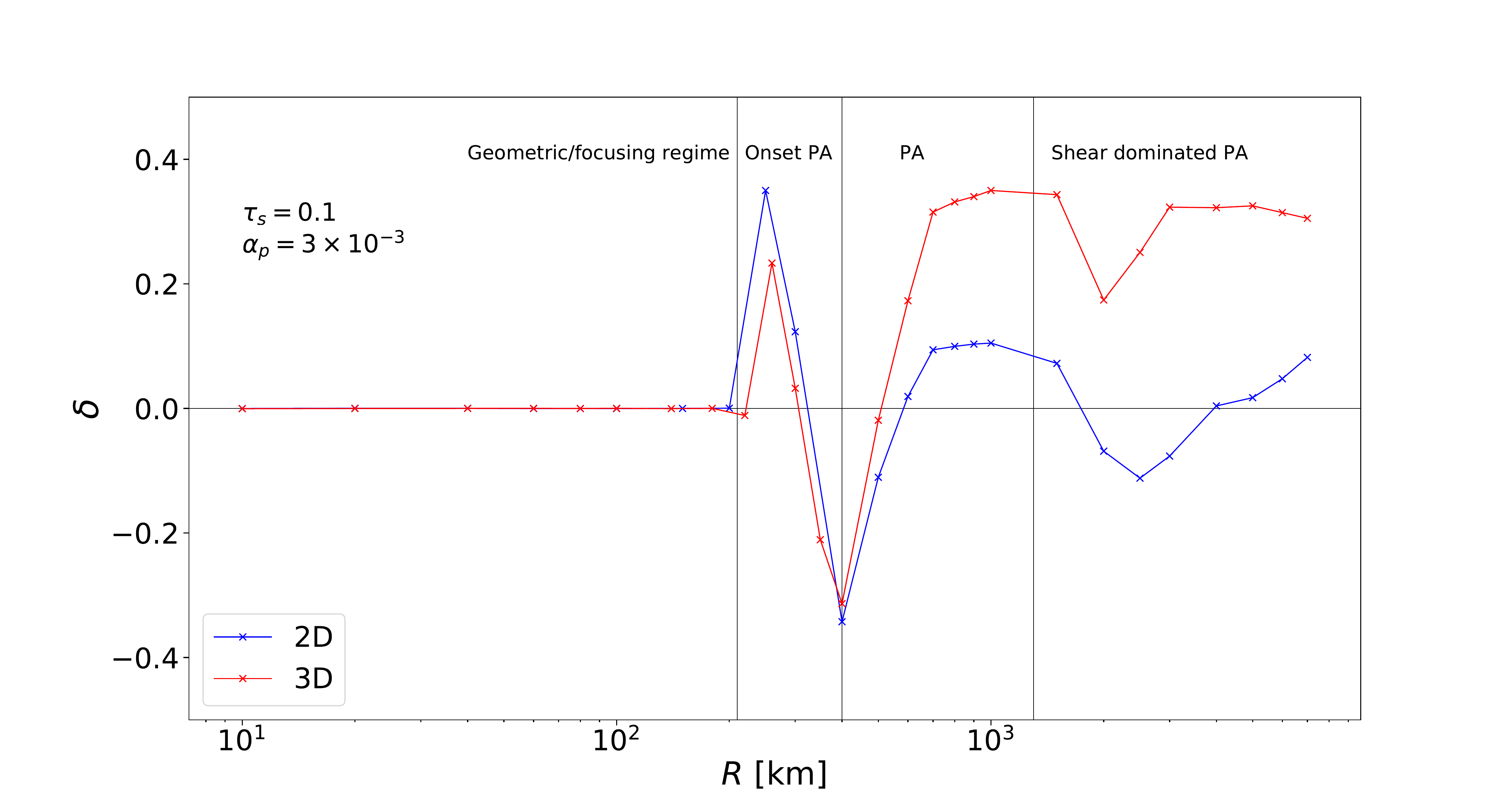}
    \caption{\textbf{The degree of asymmetry $\delta$ vs the planet radius for both 2D and 3D mean spin outcomes. }The relevant accretion regimes are subdivided by the vertical black lines.}
    \label{fig:delta}
\end{figure*}
\begin{figure*}[t]
    \centering
    \includegraphics[width=1.2\textwidth]{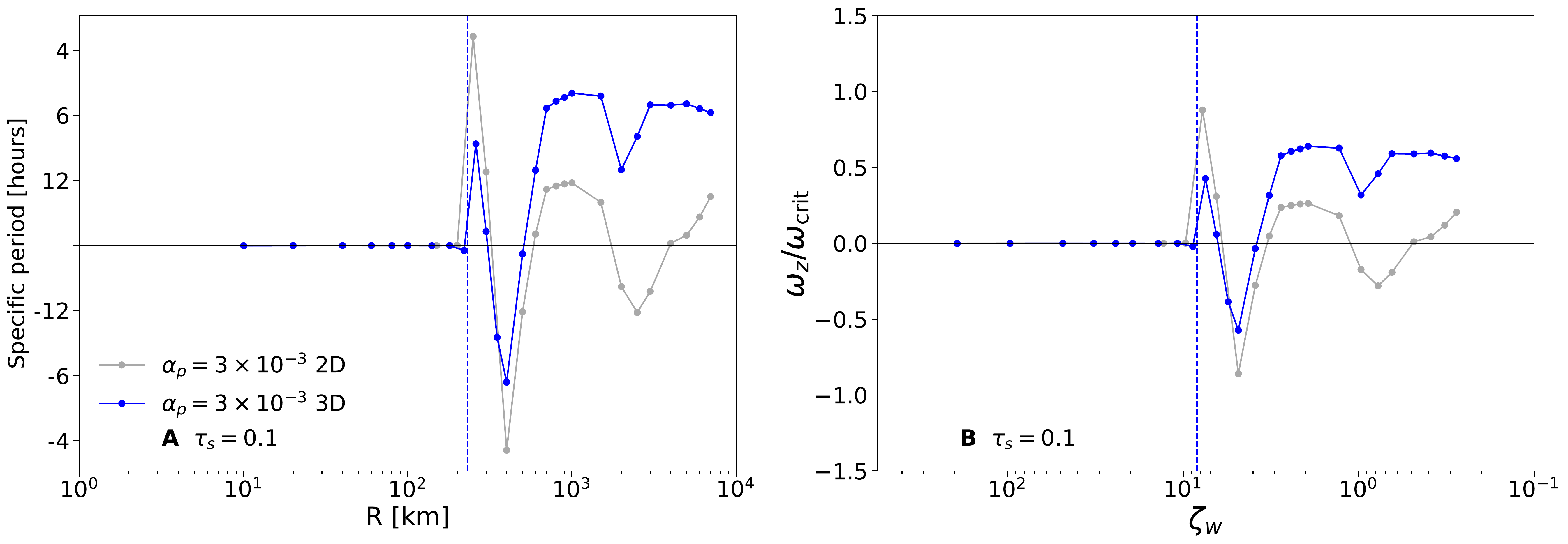}
    \caption{\textbf{Comparison of the 2D results (gray curves) with 3D results (blue curves) for the example model.} \textbf{A}: The physical quantities protoplanet radius vs specific spin period. The blue dashed vertical line indicates $R_\mathrm{PA}$. \textbf{B}:  the same data in terms of dimensionless quantities $\zeta_w$ vs the fraction of the break-up frequency.}
    \label{fig:lzgraph3d}
\end{figure*}
\begin{figure*}[t]
    \centering
    \includegraphics[width=\textwidth]{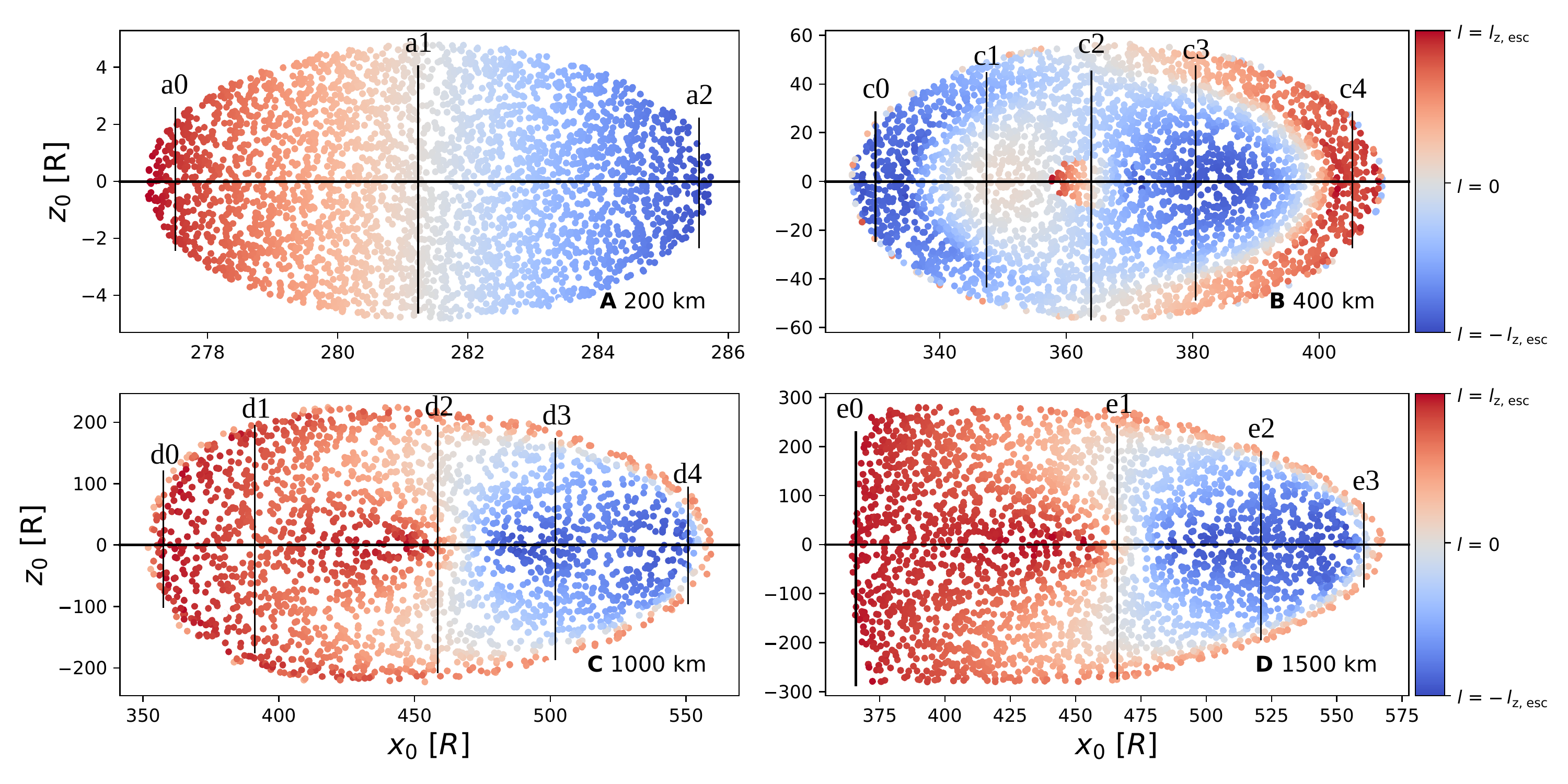}  
    \caption{\textbf{Heat maps of initial vertical and x-release coordinates for the example model, $\tau_s = 0.1$}. The color bars indicate the sign and magnitude of SAM an individual pebble imparts to the protoplanet. The vertical solid lines with unique labels indicate that the trajectory type is similar to the corresponding 2D trajectory types in Fig. \ref{fig:ballpebb} for $z_0 \neq 0$ and fixed release distance $x_0$.   For \textbf{A},  200 km; the contributions are symmetric and sum up to $\sim 0$. For \textbf{B},  400 km; the retrogade mean spin is slightly dampened with respect to the analogous 2D case.  \textbf{C},  1000 km; the Keplerian shear is equally important in the vertical release distance $z_0$, resulting in additional prograde contributions interior to the collision cross-section and dampening in retrogade contributions exterior, explaining the deformation of the elliptic shape to an egg-like shape. \textbf{D}, 1500 km; the co-rotation line interior to the protoplanet Hill sphere causes long lasting encounters and pebbles can settle vertically to the protoplanet in a timely fashion. Conversely, the exterior pebbles fail to be captured for increasing $z_0$ due to increasingly weaker protoplanet gravity, leading to the rocket-like shape.}
    \label{fig:indcont3d}
\end{figure*}
\begin{figure*}[t]
    \includegraphics[width=\textwidth]{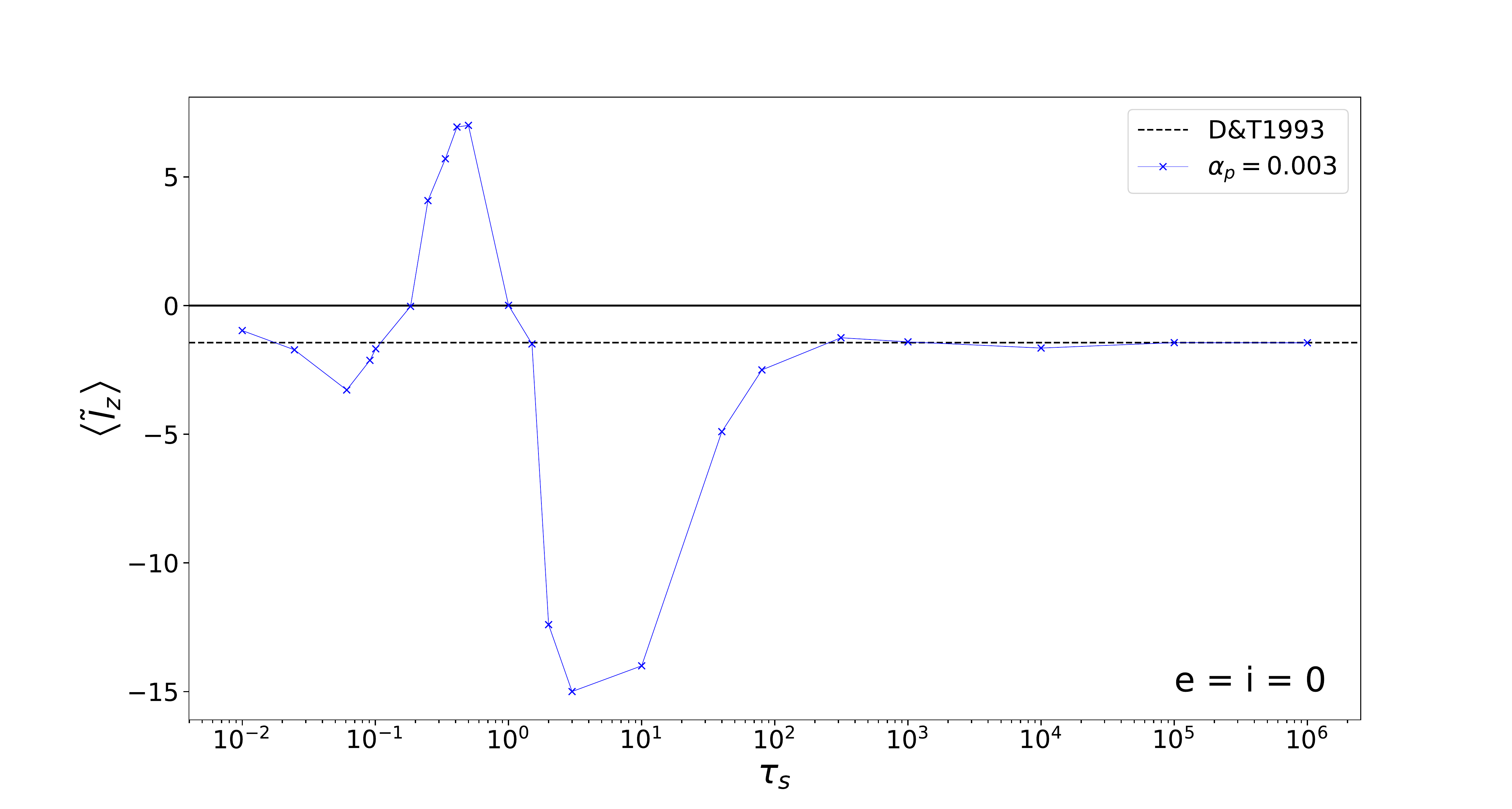} 
    \caption{\textbf{Mean dimensionless spin normalized according to \citet{DonesTremaine1993} plotted vs Stokes number.} For the convergence a fixed $\alpha_p = 0.003$ and $\zeta_w = 1$ are adopted, corresponding to the \added[]{zero dispersion $(i = e = 0)$} and high gravity regime in the gas-free case ($\tau_s > 100$) in \citet{DonesTremaine1993}.}
    \label{fig:DT}
\end{figure*}
\begin{figure}[t]
    \centering
    \includegraphics[width=.8\textwidth]{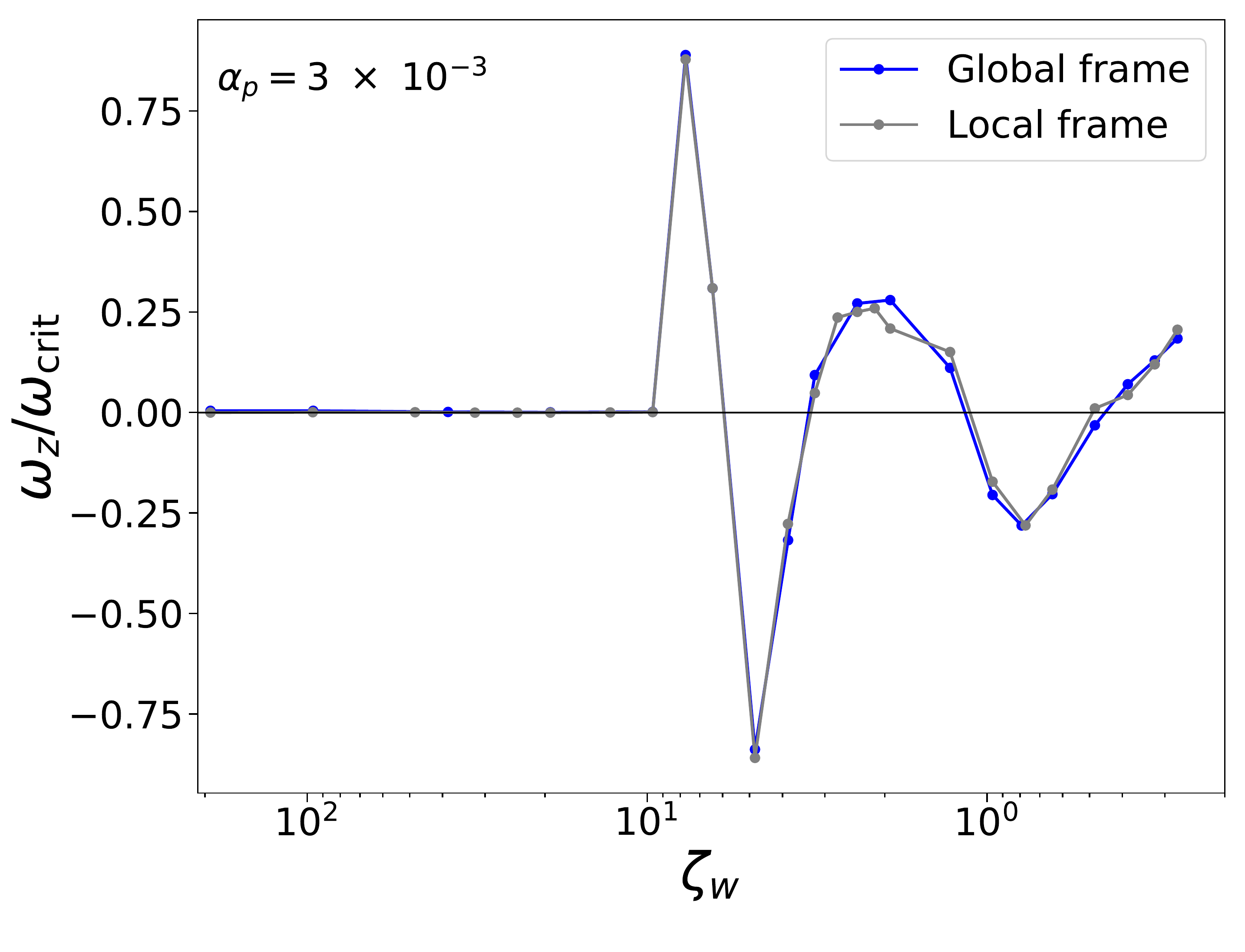}  
    \caption{\textbf{Global results vs local results.} The shear-to-headwind parameter (x-axis) vs the fraction of the break-up speed (y-axis) for $\alpha_p = 3 \times 10^{-3}$ and $\tau_s = 0.1$. The grey curve indicate results obtained from the \textit{local shearing sheet approximation}, while the blue curve is obtained from \textit{global simulations} performed with the same setup as \citet{LiuOrmel2018}.}
    \label{fig:Glvsloc}
\end{figure}
\end{document}